\documentclass[a4paper,11pt]{article}
\usepackage{jcappub} 
\usepackage{amsmath}
\usepackage{amssymb}
\usepackage{graphicx}
\usepackage{subfigure}
\usepackage{color}
\usepackage{ulem}
\usepackage{pifont}
\usepackage{makecell}
\usepackage{comment}
\begin{document}

\title{\boldmath Interpreting Pulsar Timing Array data of Gravitational Waves with Ekpyrosis-Bouncing Cosmology}

\author[a]{Taotao Qiu,}
\author[b,1]{Mian Zhu \note{Corresponding author}}

\affiliation[a]{School of Physics, Huazhong University of Science and Technology, Wuhan, 430074, China}
\affiliation[b]{Faculty of Physics, Astronomy and Applied Computer Science, Jagiellonian University, 30-348 Krakow, Poland}

\emailAdd{qiutt@hust.edu.cn}
\emailAdd{mian.zhu@uj.edu.pl}


\abstract{Recent pulsar timing array (PTA) experiments have reported strong evidence of the stochastic gravitational wave background (SGWB). If interpreted as primordial Gravitational Waves (pGWs), the signal favors a strongly blue-tilted spectrum. On the other hand, the Ekpyrosis-bouncing cosmology with a strongly blue-tilted GW spectrum, i.e., $n_T \simeq 2$, offers a potential explanation for the observed SGWB signal. In this paper, we construct a concrete Ekpyrosis-bouncing model, and show its capacity to intepret the PTA result without pathologies. Both tensor and scalar perturbations are analysed with constraints from the current observations.}

\maketitle
\flushbottom

\section{Introduction}
\label{sec:intro}

Bouncing cosmology is an important topic in the early universe cosmology \cite{Novello:2008ra}. The standard paradigm of the early universe equipped with inflation suffers from the initial singularity problem \cite{Borde:1993xh,Borde:2001nh} (see \cite{Lesnefsky:2022fen,Geshnizjani:2023edw} for recent development) and the trans-Planckian problem \cite{Brandenberger:2000wr,Martin:2000xs}. In light of that, bouncing cosmology is motivated to evade the conceptual puzzles in inflationary cosmology. Moreover, bouncing cosmology can also solve the horizon, flatness and monopole problems, as well as providing a natural explanation for the formation of the large scale structure \cite{Ijjas:2018qbo}.

Recently, pulsar timing array (PTA) collaborations, including NANOGrav \cite{NANOGrav:2023hvm,NANOGrav:2023gor}, EPTA \cite{Antoniadis:2023rey}, PPTA \cite{Reardon:2023gzh}, and CPTA \cite{Xu:2023wog}, have reported strong evidence for an isotropic stochastic GW background with a strain amplitude of order $\mathcal{O}(10^{-15})$ at the reference frequency $f = 1\, \textnormal{yr}^{-1}$. See Ref. \cite{Battista:2021rlh, Battista:2022hmv, Ashoorioon:2022raz, Ellis:2023tsl,Ellis:2023dgf,Franciolini:2023pbf,Zhang:2023lzt,Cannizzaro:2023mgc,Shen:2023pan,Du:2023qvj,Balaji:2023ehk,Zhang:2023nrs,Konoplya:2023fmh,Wu:2023hsa,Ghosh:2023aum,Huang:2023chx,Liu:2023pau,Cai:2023dls,Liu:2023ymk,Jiang:2023qbm,Cai:2023uhc,Maji:2023fhv,Addazi:2023jvg,Ellis:2023oxs,Li:2023bxy,Xiao:2023dbb,Han:2023olf,Oikonomou:2023bah,Bian:2023dnv,Huang:2023zvs,Ye:2023xyr,Guo:2023hyp,Choudhury:2023kam,Jiang:2023gfe,Oikonomou:2023bli,Choudhury:2023fwk,Lozanov:2023rcd,Oikonomou:2023qfz,Choudhury:2023fjs,Liu:2023hpw,Ye:2023tpz,Chowdhury:2023xvy,Chen:2024fir,Jiang:2024dxj,Chen:2024twp,Choudhury:2024dzw,Chen:2024mwg,Li:2024dce,Chen:2024jca, Pallis:2024mip,Li:2024oru} for interpretations of PTA results. The PTA result strongly supports a blue tensor spectrum with $n_T = 1.8 \pm 0.3$ \cite{Vagnozzi:2023lwo}. It is well-known that, the canonical inflation scenario predicts a nearly scale invariant primordial tensor spectra. \footnote{See, e.g., \cite{Wang:2014kqa} for blue tensor spectra in early universe cosmology.} On the other hand, the Ekpyrosis bouncing cosmology \cite{Khoury:2001wf,Brandenberger:2009jq}, predict a blue tensor spectra with a spectral index $2 < n_T < 3$ (see \cite{Brandenberger:2020tcr} for exceptions). Therefore, the recent PTA result might be a potential hint for the Ekpyrosis bouncing cosmology. See concrete models of Ekpyrosis bouncing cosmology in Ref. \cite{Qiu:2011cy,Cai:2012ag,Cai:2012va,Cai:2014jla,Qiu:2015nha,Wan:2015hya,Li:2016awk,Cai:2016hea,Cai:2017pga,Qiu:2018nle,Hu:2023ndc}.

In \cite{Zhu:2023lbf}, we conduct a prelimilary check on the possibility to explain the PTA data by the primordial gravitational wave (PGW) signal in non-singular cosmology. It is found that, although our toy Genesis-inflation model can potentially generate a stochastic gravitational wave background (SGWB) capable for PTA result, there are issues, such as a trans-Planckian problem and 
an oversized scalar perturbation, to be addressed in details. Therefore, it is important to study the above issue in a concrete realization.

In this paper, we will provide a concrete realization of non-singular bouncing cosmology capable with PTA observations. In bouncing scenario, the universe starts with a contraction epoch with Hubble parameter $H < 0$. After a bouncing epoch where $H$ transits from negative to positive, the universe enters into an expansion epoch. We will adopt an Ekpyrotic contraction epoch, for not only the blue tensor spectra, but also its robustness against anisotropic stress and initial conditions \cite{Ijjas:2020dws}. Unfortunately, the Ekpyrotic contraction epoch also predicts a blue-tilted scalar spectra, which is inconsistent with observations. Thus, we follow the curvaton mechanism to acquire a nearly scale-invariant scalar spectra on CMB scales, while keeping the tensor spectra blue \cite{Qiu:2013eoa}. Under the above guidance, we construct our model with a concrete action, and show that the observed PTA signals can be predicted with suitable choices of model parameters without the over-production of primordial fluctuations.

The paper is organized as follows. We present our model and the description of background evolution in Sec. \ref{sec:model}. We analyze the tensor and scalar perturbation in Sec. \ref{sec:tensorpt} and \ref{sec:scalarpt}, respectively. A numerical investigation of the tensor power spectrum and its connection to PTA observation is presented in Sec. \ref{sec:observation}. We finally conclude in Sec. \ref{sec:Conclusion}. Throughout this paper, we adopt the $(-,+,+,+)$ convention. The scalar field $\phi$ is set to be dimensionless, so the canonical kinetic term $X$, defined as $X \equiv - (1/2)g^{\mu \nu} \nabla_{\mu} \phi \nabla_{\nu} \phi$, is of dimension $[M]^2$. We use an overdot to denote the differentiation with respect to cosmic time $t$, and a prime to denote the differentiation with respect to conformal time $\tau \equiv dt/a$.

\section{Model and background}
\label{sec:model}
\subsection{Action} 

The generic action is taken to be
\begin{equation}
    \label{eq:Action}
    S = \int d^4x \sqrt{-g} \left[ M_p^2 \frac{R}{2} + K(\phi,X) + \mathcal{L}_{\textnormal{EFT}} + \mathcal{L}_{\sigma} \right] ~,
\end{equation}
where $R/2$ is the Einstein-Hilbert action. The action \eqref{eq:Action} is composed by three part, which we shall explain below. 

The k-essence action $K(\phi,X)$ is responsible for the background dynamics. We assume that the universe starts with an Ekpyrotic contraction epoch at $\phi \ll -1$, and finally evolves into an radiation-dominated expansion epoch at $\phi \gg 1$. This suggests the following asymptotic behavior of $K$:
\begin{equation}
\label{eq:Kasymp}
    \lim_{\phi \to -\infty} K(\phi,X) = X + 2V_0 e^{\sqrt{\frac{2}{q}} \phi} ~;~ \lim_{\phi \to \infty} K(\phi,X) = X^2 ~.
\end{equation}
Moreover, during the bouncing epoch, NEC is violated. We will use a ghost-condensate type action \cite{Arkani-Hamed:2003pdi}, i.e. $-X + X^2$, to violate NEC. Therefore, the k-essence action shall take the form
\begin{equation}
\label{eq:K}
    K(\phi,X) = [1-g(\phi)] M_p^2 X + \beta_2 X^2 - M_p^4 V(\phi) ~,
\end{equation}
with $g$ being the smooth top-hat function 
\begin{align}
\label{eq:g}
    g(\phi) & \nonumber = \beta_1 \left[ \frac{1 + \tanh \lambda_1(\phi - \phi_-)}{2} \right] \left[ \frac{1 + \tanh \lambda_2(\phi_+ - \phi)}{2} \right]  \\
    & + \frac{1 + \tanh \lambda_4 (\phi - \phi_+)}{2} ~,~ \phi_- < \phi_+ ~, 
\end{align}
and a potential
\begin{equation}
\label{eq:V}
    V(\phi) = -2V_0 e^{\sqrt{\frac{2}{q}} \phi}  \frac{1 - \tanh \lambda_3(\phi - \phi_-)}{2} ~,
\end{equation}
with the restriction $\lambda_3 > 1/\sqrt{2q}$. One may check that in the ansatz, \eqref{eq:K} along with \eqref{eq:g} and \eqref{eq:V}, the asymptotic behavior \eqref{eq:Kasymp} is satisfied (the potential $V(\phi)$ is exponentially suppressed so that the action is dominated by the $X^2$ term when $\phi > \phi_+$). Besides, when $\phi_- < \phi < \phi_+$, the action \eqref{eq:K} becomes 
\begin{equation}
    K(\phi,X) \simeq (1-\beta_1) M_p^2 X + \beta_2 X^2 - M_p^4 V(\phi) ~.
\end{equation}
which permits an NECV as long as $\beta_1 > 1$.

There will be ghost or gradient instability problems in the generic bouncing models within the framework of Horndeski theory \cite{Libanov:2016kfc,Kobayashi:2016xpl, Akama:2017jsa}. The simplest way to evade the instabilities is to introduce an EFT operator in the context of non-singular cosmology \cite{Cai:2016thi,Cai:2017tku,Cai:2017dyi,Li:2018ixg} (alternative approaches can be found in \cite{Kolevatov:2017voe, Mironov:2018oec,Boruah:2018pvq,Banerjee:2018svi,Qiu:2018nle,Volkova:2019jlj,Mironov:2022quk,Mironov:2022ffa,Ganz:2022zgs,Ilyas:2020qja,Ilyas:2020zcb,Zhu:2021ggm,Zhu:2021whu,Hu:2023ndc}). Since the EFT operator changes the scalar sound speed $c_s^2$ only, and has no impact on the background dynamics and tensor perturbations, we may simply use it to resolve the instability problem for bouncing cosmology, and don't need to worry about its contribution to both the background evolution and linear perturbations. 

However, the Ekpyrotic contraction period will generate a blue-tilted scalar power spectrum \cite{Creminelli:2004jg,Piao:2004jg,Piao:2004uq,Qiu:2012ia,Qiu:2012np,Shi:2021tmn}. Therefore, it is useful to introduce a curvaton field $\sigma$ coupled to the scalar field $\phi$, which can lead to the desired scale-invariant scalar power spectrum, while leaving the bouncing background unchanged \cite{Qiu:2013eoa,Lyth:2001nq,Lyth:2002my,Libanov:2011bk, Hinterbichler:2011qk,Creminelli:2011mw,Qiu:2011cy}. \footnote{Curvaton has been discussed a lot in recent years with many new models proposed, see \cite{Wang:2011dt,Feng:2013pba,Feng:2014tka,Qiu:2016mrx,Zhang:2022bde,Xiong:2024vms}.}As shown explicitly in \cite{Qiu:2013eoa}, we can adopt the follwing action for curvaton:
\begin{equation}
\label{eq:Lsigmacontraction}
    \mathcal{L}_{\sigma} = - \frac{M_p^2}{2} f^2 e^{\sqrt{2/q} \phi} (\partial \sigma)^2 ~,
\end{equation}
acquiring a scale-invariant scalar power spectrum which is constant in time. Here, $f$ is a dimensionless parameter. As is explained in the literatures cited above, the coupling function is needed so as to make the curvaton ``feel itself in an inflation-like scenario, which can give rise to the scale-invariant power spectrum required by the observations. Moreover, in \cite{Hinterbichler:2011qk, Creminelli:2011mw}, it was stated that the form of Eq. \eqref{eq:Lsigmacontraction} can be taken as an effective theory of a system with a matter field and a goldstone boson from the breakdown of conformal symmetry. Unfortunately, in the simple set \eqref{eq:Lsigmacontraction}, the curvaton field will be dominant in the expansion stage where $\phi \gg 1$. In view of that, a realistic choice of curvaton action can be
\begin{equation}
\label{eq:Lsigma}
    \mathcal{L}_{\sigma} = - \frac{1}{2} M_p^2 f^2 \frac{1 + \phi e^{\sqrt{\frac{2}{q}}\phi}}{e^{-\sqrt{\frac{2}{q}}\phi} + \phi^2 e^{\sqrt{\frac{2}{q}}\phi}} (\partial \sigma)^2 ~,
\end{equation}
which exactly has the asymptotic form \eqref{eq:Lsigmacontraction} when $\phi \ll 1$. At the expansion stage, the action \eqref{eq:Lsigma} approximates to $\mathcal{L}_{\sigma} \propto \phi^{-1} (\partial \sigma)^2$, which we shall explain in details in Sec. \ref{sec:r}. Besides, one may check that the denominator and numerator in \eqref{eq:Lsigma} is always positive. Such an inverse coupling can also be found in scalar-tensor theories such as Brans-Dicke theory \cite{Brans:1961sx}, while its fundamental interpretations still remain unexplored.

In the following, we will start with the action \eqref{eq:Action}, along with \eqref{eq:K} and \eqref{eq:Lsigma}, to pursue its background and perturbation evolutions and to see how the primordial GWs can meet with the PTA data.

\subsection{Equations of motion for background}
According to \eqref{eq:Action}, \eqref{eq:K} and \eqref{eq:Lsigma}, the Friedmann's equations are
\begin{equation}
\label{eq:Friedmann1}
    3H^2 M_p^2 = \rho = \frac{1}{2} (1-g)M_p^2 \dot{\phi}^2 + \frac{3}{4}\beta_2 \dot{\phi}^4 + M_p^4 V(\phi) + \frac{M_p^2 f^2}{2} \frac{1 + \phi e^{\sqrt{\frac{2}{q}}\phi}}{e^{-\sqrt{\frac{2}{q}}\phi} + \phi^2 e^{\sqrt{\frac{2}{q}}\phi}} \dot{\sigma}^2 ~,
\end{equation}
\begin{equation}
\label{eq:Friedmann2}
    -2\dot{H}  M_p^2 = \rho+p = (1-g)M_p^2 \dot{\phi}^2 + \beta_2 \dot{\phi}^4 + M_p^2 f^2 \frac{1 + \phi e^{\sqrt{\frac{2}{q}}\phi}}{e^{-\sqrt{\frac{2}{q}}\phi} + \phi^2 e^{\sqrt{\frac{2}{q}}\phi}} \dot{\sigma}^2 ~.
\end{equation}

Since we now have two fields, we need one additional dynamical equation. We obtain the dynamics of $\sigma$ by varying the action \eqref{eq:Action} with respect to $\sigma$:
\begin{equation}
\label{eq:sigmadynamics}
    \ddot{\sigma} + \dot{\sigma} \frac{d}{dt} \left[ \ln \left( a^3 \frac{M_p^2 f^2}{2} \frac{1 + \phi e^{\sqrt{\frac{2}{q}}\phi}}{e^{-\sqrt{\frac{2}{q}}\phi} + \phi^2 e^{\sqrt{\frac{2}{q}}\phi}} \right)  \right] = 0 ~.
\end{equation}
Apart from the constant solution $\dot{\sigma} = 0$, there's another branch of solution 
\begin{equation}
\label{eq:sigmabranch}
    a^3 M_p^2 f^2 \frac{1 + \phi e^{\sqrt{\frac{2}{q}}\phi}}{e^{-\sqrt{\frac{2}{q}}\phi} + \phi^2 e^{\sqrt{\frac{2}{q}}\phi}} \dot{\sigma} = \textnormal{const} ~.
\end{equation}

Before proceeding, we shall examine whether the curvaton field has negligible contributions to the background dynamics. In the contraction epoch, we have $L_{\sigma} \propto e^{\sqrt{2/q} \phi} (\partial \sigma)^2$, so the curvaton has no significant contribution to the background, as proved in \cite{Qiu:2013eoa}. In the radiation dominated epoch, we have $\dot{\phi} \propto t^{-1/2}$ such that $\phi$ grows as $t^{1/2}$. Thus $\dot{\sigma} \propto t^{-1}$ according to \eqref{eq:Lsigma} and the energy density of curvaton field evolves as $\rho_{\sigma} \propto \phi^{-1} \dot{\sigma}^2 = t^{-5/2}$. So $\rho_{\sigma}$ decays more rapidly than the background energy density and there is no backreaction problem in the radiation dominated epoch. In both cases, the curvaton field has no change to introduce a large back-reaction to the background.

Finally, it would be useful to write down the dynamical equation of $\phi$ in the numerical process. Since we've omitted the contributions of the curvaton field on the background level, the equation simplifies to
\begin{equation}
    \left[ (1-g)M_p^2 + 3\beta_2 \dot{\phi}^2 \right] \ddot{\phi} + 3H\dot{\phi} \left( (1-g)M_p^2 + \beta_2 \dot{\phi}^2 \right) + M_p^4 V^{\prime}(\phi) - \frac{M_p^2}{2} g^{\prime}(\phi) \dot{\phi}^2 = 0 ~.
\end{equation}

\subsection{Background evolution in different epochs}
\label{sec:bgparametrization}
In the far past $\phi \ll -1$ and $\dot{\phi} \ll M_p$ (so that the $X^2$ term is subdominate to the $X$ term), with the help of \eqref{eq:Kasymp}, the Friedmann equations \eqref{eq:Friedmann1}, \eqref{eq:Friedmann2} become
\begin{equation}
    3H^2M_p^2 = \frac{1}{2} M_p^2 \dot{\phi}^2 - 2V_0 M_p^4 e^{\sqrt{\frac{2}{q}} \phi} ~,~ -2\dot{H} = \dot{\phi}^2 ~,
\end{equation}
which gives the Ekpyrotic attractor solution 
\begin{equation}
\label{eq:bgE}
    \phi \simeq - \sqrt{\frac{q}{2}} \ln \left[ \frac{2V_0 M_p^2 t^2}{q(1-3q)} \right] ~,~ \dot{\phi} = \frac{\sqrt{2q}}{-t} ~;~ H = \frac{q}{t} < 0 ~.
\end{equation}

The effective equation-of-state (EoS) parameter in the Ekpyrotic contraction epoch is solely determined by the parameter $q$: $w_c = -1 + 2/(3q)$. The spacetime geometry evolves accordingly:
\begin{equation}
\label{eq:bgcontraction}
    a(\tau) = a_- \left( \frac{\tau_e -\tau}{\tau_{e} - \tau_-} \right)^\frac{q}{1-q} ~,~ \mathcal{H} = \frac{-q}{(1-q)(\tau_e - \tau)} < 0 ~,
\end{equation}
where $\tau \equiv \int dt/a$ is the conformal time and $\mathcal{H} = aH$ is the conformal Hubble parameter. $\tau_-$ is the end time of the contraction epoch and $a_- \equiv a(\tau_-)$ is the corresponding scale factor. Finally, $\tau_e > \tau_-$ is an integration constant. It is convenient to have the relation between $t$ and $\tau$:
\begin{equation}
\label{eq:ttau}
    t = \int a(\tau)d\tau = -a_- (1-q) \left[ \frac{(\tau_e-\tau)}{ (\tau_e - \tau_-)^q } \right]^{\frac{1}{1-q}} < 0 ~,
\end{equation}
up to an integration constant. 

The evolution of fluctuations in bouncing epoch is generically involved. Fortunately, the duration of the bouncing epoch is usually taken to be short. For example, in \cite{Xue:2011nw} it is pointed out that the sourced anisotropy in the bouncing epoch grows exponentially. Thus a short bounce is favored for the model to be free from anisotropic stress problem. In this case, the bouncing epoch can be parametrized in the following (see e.g., Ref. \cite{Cai:2007zv,Cai:2008qw} for details)
\begin{equation}
\label{eq:bouncegeo}
    \mathcal{H} \simeq \alpha^2 (\tau - \tau_B) ~,~ a = a_B e^{\frac{1}{2} \alpha^2 (\tau - \tau_B)^2} ~,~ \tau_- < \tau < \tau_+ ~,
\end{equation}
where $\tau_B$ is an integration constant and $\tau_+$ is the end of bouncing epoch (or equivalently, the beginning of expanding epoch). The validation of the linear approximation requires $\vert \alpha (\tau-\tau_B)\vert < 1$.

While the parameterization \eqref{eq:bouncegeo} can be understood as a taylor expansion around the bouncing point, the dynamics of scalar field $\phi$ in bouncing phase is more involved. In \cite{Cai:2012va} the authors present an approximated formulae
\begin{equation}
\label{eq:bouncephi}
    \dot{\phi} \simeq M_p \sqrt{\frac{2(\beta_1 - 1)}{3\beta_2}} e^{- t^2/T^2} ~,
\end{equation}
with $T$ a free parameter, which roughly equals to a quarter of the duration of bouncing phase. The validation of \eqref{eq:bouncephi} is examined by the authors in \cite{Cai:2012va} by numerics. We will use \eqref{eq:bouncephi} to describe the dynamics of $\phi$ in bouncing phase in the rest of the paper.

After the bouncing epoch, the universe enters into an expansion phase. The Lagrangian of $\phi$ is $K(\phi,X) = X^2$, so that $\rho = 3X^2$ and $p = X^2$ and the Equation-of-state parameter for $\phi$ becomes $w_{\phi} = 1/3$. The universe will then be radiation-dominated and the creation of a thermal bath can be realized by the decaying from the scalar field $\phi$. The corresponding background dynamics becomes
\begin{equation}
    a = a_+ \frac{\tau - \tilde{\tau}}{\tau_+ - \tilde{\tau}} ~,~ \mathcal{H} = \frac{1}{\tau - \tilde{\tau}} ~.
\end{equation}

\subsection{Matching condition}
In order to maintain the continuity of $a$ and ${\cal H}$, the background dynamics in each epoch can be furtherly matched by the junction condition at the transition surface $\tau = \tau_-$ and $\tau = \tau_+$. Firstly, we set the bouncing point $H=0$ to be the zero point of $\tau$, (i.e., $\tau_B = 0$), so that $\tau_- < 0$, $\tau_+ > 0$ and
\begin{equation}
\label{eq:bgbounce}
    \mathcal{H} = \alpha^2 \tau ~,~ a = a_B e^{\frac{1}{2} \alpha^2 \tau^2} ~,~ \tau_- < \tau < \tau_+ ~,
\end{equation}
the continuity at $\tau = \tau_-$ gives
\begin{equation}
    \tau_e = \tau_- - \frac{q}{(1-q)\alpha^2 \tau_-} ~,
\end{equation}
and the continuity at $\tau = \tau_+$ gives $\tilde{\tau} = \tau_+ - (\alpha^2 \tau_+)^{-1}$. For the sake of simplicity, let's work in a symmetric bounce, i.e., $|\tau_-| = \tau_+$. Then we have $a_- = a_+ = a_B e^{\alpha^2 \tau_+^2/2}$. Implemented with those conditions, we summarize the evolution of scale factor
\begin{equation}
\label{eq:abg}
a(\tau) = \left\{ \begin{array}{cl}
  & a_B e^{ \frac{1}{2} \alpha^2 \tau_+^2} \left[ 1 + \frac{1-q}{q} \alpha^2 |\tau_-| (\tau_- - \tau) \right]^\frac{q}{1-q} ~,~ \tau < \tau_- ~, \\ \\
  & a_B e^{\frac{1}{2} \alpha^2 \tau^2} ~,~ \tau_- < \tau < \tau_+ ~, \\ \\
  & a_B e^{ \frac{1}{2} \alpha^2 \tau_+^2} \left[ 1 + \mathcal{H}_+ (\tau - \tau_+) \right] ~,~ \tau > \tau_+ ~.
\end{array} \right.
\end{equation}
and of conformal Hubble parameter
\begin{equation}
\label{eq:cHbg}
\mathcal{H}(\tau) = \left\{ \begin{array}{cl}
  & \alpha^2 \tau_- \left[ 1 + \frac{1-q}{q} \alpha^2 |\tau_-| (\tau_- - \tau) \right]^{-1} ~,~ \tau < \tau_- ~, \\ \\
  & \alpha^2 \tau ~,~ \tau_- < \tau < \tau_+ ~, \\ \\
  & \alpha^2 \tau_+ \left[ 1 + \alpha^2 \tau_+ (\tau - \tau_+) \right]^{-1} ~,~ \tau > \tau_+ ~.
\end{array} \right.
\end{equation}
The background value is thus fixed by four parameters: $\alpha$, $\tau_+$, $a_B$ and $q$.

\begin{figure}[ht]
    \centering
    \includegraphics[width=0.4\linewidth]{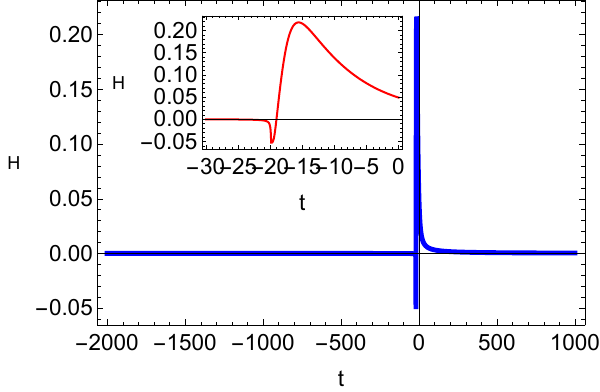}
    \includegraphics[width=0.45\linewidth]{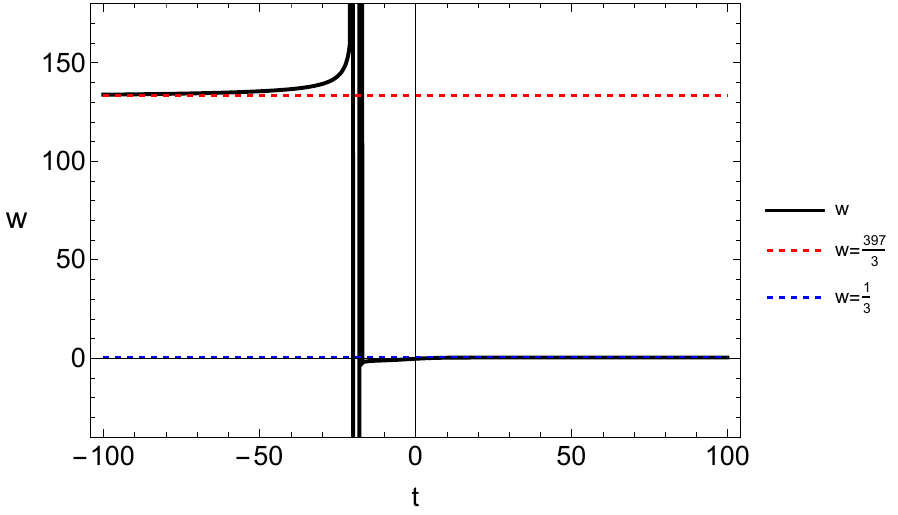}
    \caption{Sketch of Hubble parameter and Equation-of-state parameter near the bouncing phase.}
    \label{fig:Ht}
\end{figure}

We numerically evaluate the background dynamics using the following parameter set $\beta_1 = 1.1$, $\beta_2 = 1$, $\phi_- = -0.1$, $\phi_+ = 3.7$, $V_0 = 10^{-1}$, $q = 0.005$, and present the evolution of Hubble parameters in Fig. \ref{fig:Ht}. In the figure, the Hubble parameter goes from below zero to above zero, indicating that bounce indeed occurs. While before the bounce the EoS is very large as in the Ekpyrosis phase $w = -1+2/(3q)$, after the bounce the universe enters a radiation-dominated era where the EoS is nearly $1/3$. Note that all model parameters are normalized to be dimensionless.

\section{Tensor perturbations}
\label{sec:tensorpt}
\subsection{Generic formalism}
Now we turn to the tensor perturbations generated in our model. The quadratic action of tensor perturbation is
\begin{equation}
	S_{2,T} = \int d\tau d^3x \frac{a^2}{8} M_p^2 \left[ \gamma_{ij}^{\prime 2} - c_T^2 (\partial \gamma_{ij})^2 \right] ~,
\end{equation}
where $\gamma_{ij}$ represents the tensor perturbation which is dimensionless. Here for simplicity, we don't distinguish the polarization modes for the tensor sector. Since the matter sector is minimally coupled to gravity, the propogation speed of GWs is $c_T^2 = 1$. The dynamical equation is accordingly
\begin{equation}
    \label{eq:tensoreq}
    \nu_k^{\prime \prime} + \left(k^2 - \frac{a^{\prime \prime}}{a} \right) \nu_k = 0 ~,
\end{equation}
where $\nu_k \equiv a\gamma_k/(2\sqrt{M_p})$ is the mode function of tensor perturbation.

We will need the initial condition to solve \eqref{eq:tensoreq}. The universe starts in an Ekpytoric contraction configuration described by \eqref{eq:bgE} and \eqref{eq:bgcontraction}, during which the dynamical equation is:
\begin{equation}
    \nu_k^{\prime \prime} + \left[ k^2 - \frac{q(2q-1)}{(1-q)^2 (\tau_e - \tau)^2} \right] \nu_k = 0 ~,
\end{equation}
whose general solution is the Hankel function $H_{\nu}^{(1)} [k(\tau_e - \tau)]$ and $H_{\nu}^{(2)} [k(\tau_e - \tau)]$. Imposing the vacuum initial condition $\nu_k \sim e^{-ik\tau}/\sqrt{2k}$, the mode function evolves as
\begin{equation}
\label{eq:nue}
    \nu_k(\tau) \simeq \frac{\sqrt{\pi (\tau_e - \tau)}}{2} H_{\nu}^{(1)} [k(\tau_e - \tau)] ~,~ \nu \equiv \frac{1-3q}{2(1-q)} ~.
\end{equation}
Note that $\nu_k$ has a dimension of $[M]^{-1/2}$, as expected from the definition. We will set Equation \eqref{eq:nue} as the initial condition. The dimensionless primordial tensor power spectrum is defined as 
\begin{equation}
    P_T = 2 \cdot \frac{k^3}{2\pi^2 M_p^3} |\gamma_k|^2 = \frac{4k^3}{\pi^2 M_p^2} \left| \frac{\nu_k}{a} \right|^2 ~,
\end{equation}
where the factor $2$ comes from the two tensorial polarizations.

\subsection{Dynamics of tensor fluctuations}
Our task is to evaluate the tensor power spectrum using \eqref{eq:tensoreq} and \eqref{eq:nue} for $k < \mathcal{H}_+$ modes. As we argued above, the modes of interest are super-horizon at the beginning of radiation dominated epoch, $\tau = \tau_+$. These modes are conserved before the horizon re-entry, so it suffices to evaluate the tensor spectrum at $\tau = \tau_+$. The dynamics of tensor perturbation during the contraction phase is dictated by \eqref{eq:nue}. On super-horizon scale, the tensor perturbation and its derivative at $\tau = \tau_-$ is
\begin{equation}
    \nu_k (\tau_-) = -\frac{i\sqrt{\tau_e - \tau_-}}{2\sqrt{\pi}} \Gamma(\nu) \left[ \frac{2}{k(\tau_e - \tau_-)} \right]^{\nu} = -\frac{2^{\nu-1} i}{\sqrt{\pi}} \Gamma(\nu) k^{-\nu} \left[ \frac{\alpha^2 (1-q) \tau_+}{q} \right]^{\nu-\frac{1}{2}} ~,
\end{equation}
\begin{equation}
    \nu_k^{\prime} (\tau_-) = \frac{i \Gamma(\nu) 2^{\nu}}{(1-2\nu) \sqrt{\pi}} k^{-\nu} \left[ \frac{\alpha^2 (1-q) \tau_+}{q} \right]^{\nu+\frac{1}{2}} ~,
\end{equation}
where we use the fact $|\tau_-| = -\tau_- = \tau_+$. 

With \eqref{eq:bgbounce}, the dynamical equation for tensor perturbation in the bouncing phase approximates to 
\begin{equation}
\label{eq:nukb}
    \nu_k^{\prime \prime} + \left( k^2 - \alpha^2 \right) \nu_k = 0 ~,
\end{equation}
whose general solution is 
\begin{equation}
\label{eq:nuksolbounce}
    \nu_k = b_{T,1} e^{\sqrt{\alpha^2 - k^2} \tau} + b_{T,2} e^{-\sqrt{\alpha^2 - k^2} \tau} ~,~ \tau_- < \tau < \tau_+ ~.
\end{equation}
We can interpret \eqref{eq:nukb} in the following way. Modes with $k^2 < \alpha^2$ will experience a tachyonic instability \cite{Qiu:2011cy,Easson:2011zy}, where the corresponding fluctuations are amplified. On the other hand, modes with $k^2 > \alpha^2$ will simply exhibit an oscillatory behavior.

Finally, in the radiation dominated epoch, we have $a^{\prime \prime}/a = 0$, thus all modes of our interest oscillate in this epoch, and their amplitude remain fixed. The power spectrum in radiation dominated epoch is invariant, and it would be sufficient for us to evaluate $P_T$ at $\tau = \tau_+$.

\subsection{Horizon crossing in bouncing cosmology}
\label{sec:horizon}
\begin{figure}[ht]
    \centering
    \includegraphics[width=0.8\linewidth]{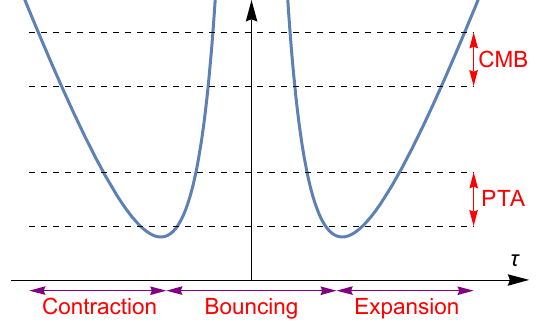}
    \caption{The evolution of Hubble horizon in bouncing cosmology.}
    \label{fig:horizon}
\end{figure}

We depict the evolution of Hubble horizon, $|H|^{-1}$, in Fig. \ref{fig:horizon}. It is easy to see that only modes with $k \leq |\mathcal{H}_-| = \mathcal{H}_+$ has chances to be super-horizon in the bouncing scenario. Those modes will cross the horizon firstly in the contraction phase, and re-enter then re-exit the horizon during the bouncing phase, which finally re-enter the horizon in the radiation dominated phase. Modes with $k > \mathcal{H}_+$ will be sub-horizon in the whole cosmic evolution. 

In fact, the cross of the Hubble horizon of the primordial fluctuations is equivalent to its classicalization, which is crucial for them to be observable. Modes which never cross the Hubble horizon has a spurious divergence thus need specific treatment. As we are interested in the intepretation of recent PTA signals from PGWs originated from the contraction phase, we shall terminate our power spectrum at the scale $k_{\ast} = \mathcal{H}_+$, where $k_\ast$ refers to the upper limit of the PTA detectability. Moreover, for range of smaller $k$ (larger scales), the CMB constraint must be satisfied as well. 

\subsection{Junction conditions and power spectra}
\label{sec:PTformulae}
The remaining task is to evaluate the power spectra for the modes with $k \leq \mathcal{H}_+ = \alpha^2 \tau_+$. Here, $\alpha \tau_+$ should be at most of $\mathcal{O}(1)$ order, otherwise the scale factor shall change intensively in the bouncing epoch, in constrast with our short bounce assumption. Therefore, we have $\mathcal{H}_+ \leq \alpha$, thus modes of our interest shall satisfy the condition $k < \alpha$. As a result, these modes will experience a tachyonic growth in the bouncing epoch as suggested by \eqref{eq:nuksolbounce}. In this case, the exponential growing mode will dominate over the exponential dacaying mode, so it suffices to evaluate $b_{T,1}$.

We calculate $b_{T,1}$ by applying the junction condition for tensor fluctuations, i.e., the mode function and its first derivative is continuous on the transition surface $\tau = \tau_-$:
\begin{equation}
    -\frac{2^{\nu-1} i}{\sqrt{\pi}} \Gamma(\nu) k^{-\nu} \left[ \frac{\alpha^2 (1-q) \tau_+}{q} \right]^{\nu-\frac{1}{2}} = b_{T,1} e^{\sqrt{\alpha^2 - k^2} \tau_-} + b_{T,2} e^{-\sqrt{\alpha^2 - k^2} \tau_-} ~,
\end{equation}
\begin{equation}
    \frac{i \Gamma(\nu) 2^{\nu}}{(1-2\nu) \sqrt{\pi}} k^{-\nu} \left[ \frac{\alpha^2 (1-q) \tau_+}{q} \right]^{\nu+\frac{1}{2}} = \sqrt{\alpha^2 - k^2} \left( b_{T,1} e^{\sqrt{\alpha^2 - k^2} \tau_-} - b_{T,2} e^{-\sqrt{\alpha^2 - k^2} \tau_-} \right) ~.
\end{equation}
After some calculation one finds that it leads to
\begin{equation}
    b_{T,1} = e^{-\sqrt{\alpha^2 - k^2} \tau_-} \frac{2^{\nu-2} i}{\sqrt{\pi}} \Gamma(\nu) k^{-\nu} \left[ \frac{q}{\alpha^2 (1-q) \tau_+} \right]^{\frac{q}{1-q}} \left[ \frac{\alpha^2 (1-q)^2 \tau_+}{\sqrt{\alpha^2 - k^2} q^2} - 1 \right] ~,
\end{equation}
where we apply $1 - 2\nu = 2q/(1-q)$.

The tensor power spectrum at $\tau = \tau_+$ is thus
\begin{equation}
    P_T = 2^{-\frac{1+q}{1-q}} \frac{\Gamma^2(\nu)H_+^2}{\pi^3M_p^2 a(\tau_+)^2} k^{\frac{2}{1-q}} e^{2\sqrt{\alpha^2 - k^2} \tau_+} \left[ \frac{q}{\alpha^2 (1-q) \tau_+} \right]^{\frac{2q}{1-q}} \left[ \frac{\alpha^2 (1-q)^2 \tau_+}{\sqrt{\alpha^2 - k^2} q^2} - 1 \right]^2 ~.
\end{equation}
In terms of Hubble parameters:
\begin{equation}
    P_T = 2^{-\frac{1+q}{1-q}} \frac{\Gamma^2(\nu)}{\pi^3M_p^2} e^{\frac{2\sqrt{\alpha^2 - k^2} \mathcal{H}_+}{\alpha^2}} \left( \frac{q}{1-q} \right)^{\frac{2q}{1-q}} \left( \frac{k/a_B}{H_+} \right)^{\frac{2}{1-q}} \left[ \frac{(1-q)^2 \mathcal{H}_+}{\sqrt{\alpha^2 - k^2} q^2} - 1 \right]^2 ~.
\end{equation}

We have $H_+ = \alpha^2 t_+$, since $\alpha t_+$ can be at most $\mathcal{O}(1)$, $H_+$ cannot exceed $\alpha$ too much. As a result, $\mathcal{H}_+ \simeq a_B H_+$ must be much smaller than $\alpha$, since $a_B \ll a_{\rm today}$. Similarly, we have $k \leq k_{\ast} = \mathcal{H}_+ \ll \alpha$. Thus, unless we take a vanishing $q$, the terms in bracket will approximates to unity. The expression simplifies to 
\begin{equation}
\label{eq:PT}
    P_T \simeq 2^{-\frac{1+q}{1-q}} \frac{\Gamma^2(\nu)}{\pi^3} \left( \frac{H_+}{M_p} \right)^2 \left( \frac{q}{1-q} \right)^{\frac{2q}{1-q}} \left( \frac{k/a_B}{H_+} \right)^{\frac{2}{1-q}} ~.
\end{equation}
It's easy to see that the tensor power spectrum is governed by three parameters: the Hubble parameter at the end of bouncing phase $H_+$, the model parameter $q$ which essentially determine the scale dependence of the primordial fluctuations, and the scale factor $a_B$ which is relevant to the post-bouncing phases. 

From the above results we can see that, for $q\lesssim 1$, $2/(1-q)>0$, and we can get a blue-tilted tensor power spectrum. For current PTA data, it requires $n_T = 1.8 \pm 0.3$ \cite{Vagnozzi:2023lwo}, which suggests $q < 0.0476$. 

Moreover, in the minimal setup, the spectral energy density parameter $\Omega_{GW} (k)$ is related to the
primordial tensor spectrum, defined as the energy density of the
GWs per unit logarithmic frequency, by
\begin{equation}
    \Omega_{GW} (k) \equiv \frac{1}{3H^2} \frac{d \rho_{\rm GW}}{d\ln k} \simeq 10^{-6} P_T(k) ~.
\end{equation}

\section{Parameter Choices against Constraints}
\label{sec:observation}

The above results shall be confronted to several theoretical and observational constraints, which can be used to confine model parameters. 

First, we discuss trans-Planckian problem, i.e., the physical wavelength of primordial fluctuations we concerned here cannot be
smaller than the Planck length. The power spectrum is terminated at the scale $k_{\ast}$, so the scenario is free from trans-Planckian problem if
\begin{equation}
    \lambda_{\ast} = \frac{1}{(k_{\ast}/a_B)} > l_P ~\to~ \frac{H_+}{M_p} < 1 ~,
\end{equation}
since at $a=a_B$ the physical wavelength takes its minimum. Here we use the fact $k_{\ast} \simeq \mathcal{H}_+(\tau_+) = a_B H_+$.

Next we are ready to confront our model with observations. As is shown in \cite{NANOGrav:2023hvm, Vagnozzi:2023lwo}, the tensor power spectrum at the PTA frequency range ($f = 10{\rm nHz}$) is given by $P_T (f = 10{\rm nHz}) \sim 10^{-3}$. On the other hand, it is useful to define the effective e-folding number counting from $\tau_+$ to today:
\begin{equation}
    N \equiv \ln \left( \frac{a_{\rm today}}{a_B} \right) ~,
\end{equation}
then the tensor power spectrum \eqref{eq:PT} expressed by $N$ becomes
\begin{equation}
    P_T(\tau_+) \simeq 2^{-\frac{1+q}{1-q}} \frac{\Gamma^2(\nu)}{\pi^3} \left( \frac{H_+}{M_p} \right)^2 \left( \frac{q}{1-q} \right)^{\frac{2q}{1-q}} a_{\rm today} \left( \frac{k/a_{\rm today}}{H_+} \right)^{\frac{2}{1-q}} e^{\frac{2}{1-q} N} ~.
\end{equation}
Here, we are evaluating the tensor fluctuation at $\tau_+$, since after that, the universe begins expansion and primordial tensor fluctuations on super-horizon scales are frozen and directly connected to observations. We also use the approximation $a(\tau_+) \simeq a_B$. Connecting with observations where $a_{\rm today} = 1$ and $k_{\rm PTA}/a_{\rm today} = 10^6 \rm{Mpc}^{-1} = 6.6 \times 10^{-33} {\rm GeV}$, the formula becomes
\begin{align}
    P_{T}(k_{\rm PTA},\tau_+) & \nonumber = 2^{-\frac{1+q}{1-q}} \left( \frac{q}{1-q} \right)^{\frac{2q}{1-q}} \frac{\Gamma^2(\nu)}{\pi^3}  e^{\frac{2}{1-q} N} \\
    & \times \left( \frac{H_+}{M_p} \right)^2 \left( \frac{5.5 \times 10^{-52} M_p}{H_+} \right)^{\frac{2}{1-q}} ~,
\end{align}
the value of which should be determined by model parameters.

We conclude the model parameters as well as the observables in Table. \ref{tab:result}. Notice that, the result is insensitive to $\lambda$'s as long as $\lambda \gg 1$, as they only decide the smoothness of top-hat functions, and we choose $\lambda_i = 20$, $i = 1,2,3,4$. Also, the parameters $\phi_-$, $\phi_+$ determines the start and end of the bouncing phase. Here we're dedicately design the value of them to let $|H_-| \simeq H_+$ to make the analytical investigation simpler \footnote{ Of course, this assumption is not mandatory for a realistic model. }. Finally, we haven't made any assumptions on the late-time evolution of the universe, so the parameter $N$ remains free in our model. We choose the value of $N$ such that $P_T(10 \rm{nHz}) = 10^{-3}$.  
\begin{table}[h]
    \centering
    \begin{tabular}{|lllllll|l|ll|}
\hline
\multicolumn{7}{|c|}{Model parameters}  & \multicolumn{1}{c|}{Variable} & \multicolumn{2}{c|}{Observables}   \\ \hline
\multicolumn{1}{|l|}{$\beta_1$} & \multicolumn{1}{l|}{$\beta_2$} & \multicolumn{1}{l|}{$\phi_-$} & \multicolumn{1}{l|}{$\phi_+$} & \multicolumn{1}{c|}{$V_0$} & \multicolumn{1}{c|}{$q$} & \multicolumn{1}{c|}{$N$} & \multicolumn{1}{c|}{$H_+$} & \multicolumn{1}{l|}{$P_T$(10nHz)} & $n_T$ \\ 
\hline
\multicolumn{1}{|l|}{1.1}         & \multicolumn{1}{l|}{1}                & \multicolumn{1}{l|}{$0.0$}                        & \multicolumn{1}{l|}{$2.8$}                       & \multicolumn{1}{l|}{$10^{-2}$}                      & \multicolumn{1}{l|}{$0.002$}  & \multicolumn{1}{l|}{115.97}     &      \multicolumn{1}{c|}{$0.20$}                 & \multicolumn{1}{l|}{$1.0 \times 10^{-3}$}            &  \multicolumn{1}{l|}{2.00}    \\ 
\hline
\multicolumn{1}{|l|}{1.1}     & \multicolumn{1}{l|}{1}                    & \multicolumn{1}{l|}{$-0.1$}                        & \multicolumn{1}{l|}{$3.7$}                       & \multicolumn{1}{l|}{$10^{-1}$}                       & \multicolumn{1}{l|}{0.005}     & \multicolumn{1}{l|}{116.06}  & \multicolumn{1}{c|}{$0.22$}     &  \multicolumn{1}{l|}{$1.0 \times 10^{-3}$}  &  \multicolumn{1}{l|}{2.01}    \\
\hline
\multicolumn{1}{|l|}{1.1} & \multicolumn{1}{l|}{1} & \multicolumn{1}{l|}{$-0.1$} & \multicolumn{1}{l|}{$2.8$}  & \multicolumn{1}{l|}{$10^{-2}$} & \multicolumn{1}{l|}{0.01}  & \multicolumn{1}{l|}{$115.96$} &   \multicolumn{1}{c|}{$0.20$} &   \multicolumn{1}{l|}{$1.0 \times 10^{-3}$} &   \multicolumn{1}{l|}{2.02}   \\ 
\hline
\multicolumn{1}{|l|}{1.1} & \multicolumn{1}{l|}{1} & \multicolumn{1}{l|}{$0.0$} & \multicolumn{1}{l|}{$3.6$}  & \multicolumn{1}{l|}{$10^{-3}$} & \multicolumn{1}{l|}{0.02}        & \multicolumn{1}{l|}{116.18}            &  \multicolumn{1}{c|}{$0.25$}   &   \multicolumn{1}{l|}{$1.0 \times 10^{-3}$} &   \multicolumn{1}{l|}{2.04}  \\ 
 \hline
\end{tabular}
    \caption{The primordial tensor power spectrum for different model parameters. Here we normalize the parameters by setting $M_p=1$.}
    \label{tab:result}
\end{table}

\begin{figure}[ht]
    \centering
    \includegraphics[width=0.8\linewidth]{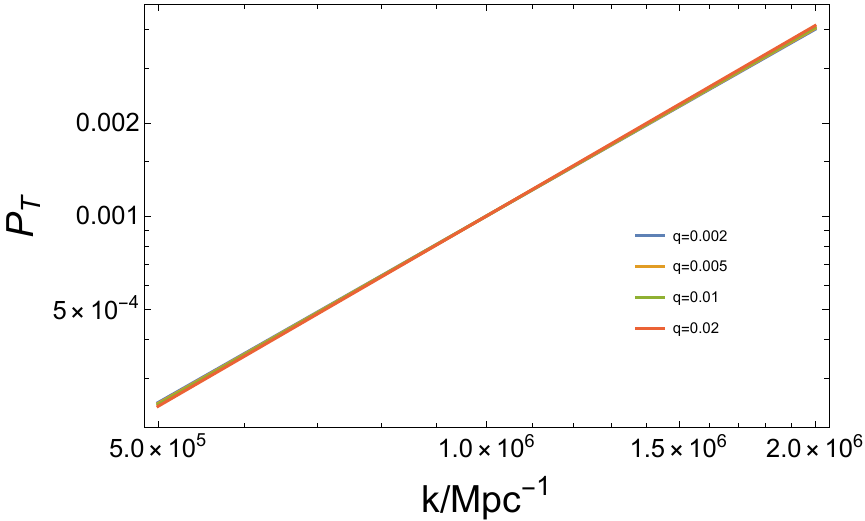}
    \includegraphics[width=0.8\linewidth]{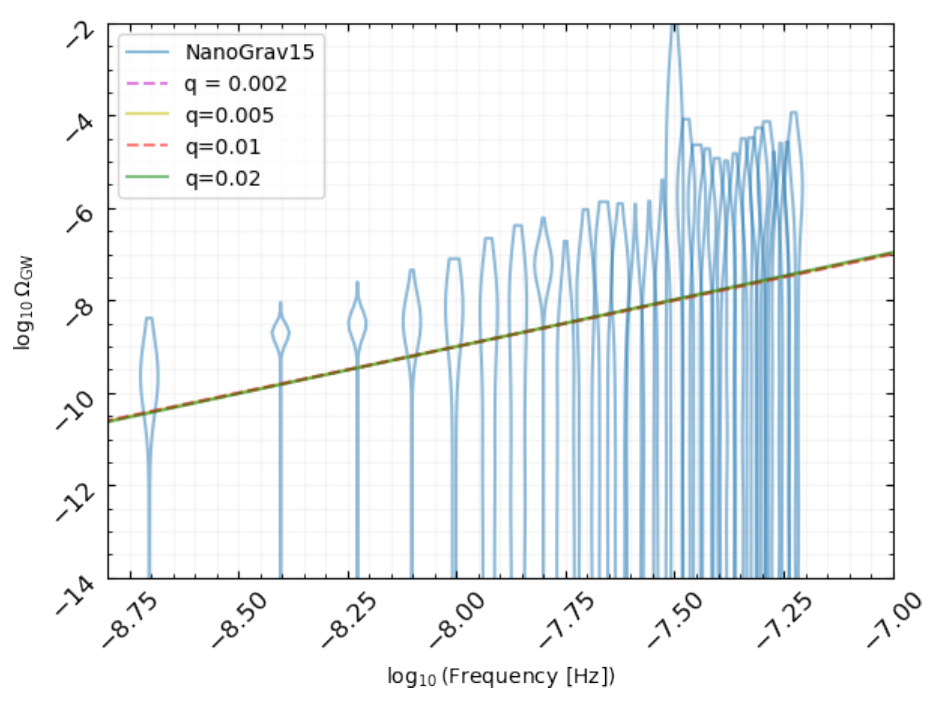}
    \caption{The tensor power spectrum and energy density spectrum for different value of $q$.}
    \label{fig:observation}
\end{figure}

We present the resultant primordial tensor power spectrum for different model parameters in Table. \ref{tab:result}. It's easy to see that the PTA result can be interpreted with certain choice of model parameters. Utilizing the parameters in Table. \ref{tab:result}, We numerically plot the results of $P_T$ and $\log_{10}\Omega_{GW}$ in Fig. \ref{fig:observation}. In the plot, we make comparison of the ${\Omega_{GW}}$ curves to the  NANOGrav 15-year data set and find that the curves fit the data very well. Moreover one can observe that, although the paramter $q$ spans over {\cal O}(10) in the sets of parameters, the curves get very closed to each other, which means that the results is not quite sensitive on the parameter $q$.

Finally, let's determine the maximum wavenumber $k_\ast$ below which we're concerning. From the above result, one can parameterize the tensor spectrum as a simpler form:
\begin{equation}
    P_T(k)\simeq 10^{-3}\left(\frac{k}{10^6 \rm{Mpc}^{-1}}\right)^{\frac{2}{1-q}}~,
\end{equation}
and in order to make the perturbative treatment work, one must have $P_T<1$ for $k\leq k_\ast$. This give rise to
\begin{equation}
\label{eq:constraintr}
    \ln \left( \frac{k_{\ast}}{10^6 {\rm Mpc}^{-1}} \right) < \frac{1-q}{2} \ln 10^3 = 3.5(1-q) ~.
\end{equation}
As the observed PTA data prefers $q \leq 0.0476$, the value $k_{\ast}$ can be at most of $\mathcal{O}(10^7) {\rm Mpc}^{-1}$ order.

\section{Scalar perturbations}
\label{sec:scalarpt}
\subsection{Scalar perturbation from the Ekpyrotic field $\phi$}

The dynamics for the fluctuation of the Ekpyrotic field $\phi$ is \cite{Zhu:2021whu}
\begin{equation}
    \mu_k^{\prime \prime} + \left( k^2 - \frac{z_s^{\prime \prime}}{z_s} \right) \mu_k = 0 ~,
\end{equation}
where $\mu_k \equiv z_s \delta \phi$ is the mode function for $\phi$ and 
\begin{equation}
\label{eq:zsgeneral}
    \frac{z_s^2}{a^2} \equiv \frac{\dot{\phi}^2 K_X + \dot{\phi}^4 K_{XX}}{H^2 M_p^2} = \frac{3\beta_2 \dot{\phi}^4 + 2(1-\beta_1) M_p^2 \dot{\phi}^2}{4M_p^2H^2} ~.
\end{equation}

As we discussed in Sec. \ref{sec:horizon}, only the modes that becomes super-horizon in the contraction phase are of observational interest to us. For these modes, the power spectrum from $\phi$ at $\tau = \tau_-$ is evaluated as \cite{Zhu:2021whu}
\begin{equation}
    P_{\phi}(k,\tau_-) \simeq \frac{\Gamma^2(\nu) (1-2\nu)^{1-2\nu}}{3\pi^3 2^{8-4\nu}} \left( \frac{k}{|\mathcal{H}_-|}\right)^{\frac{2}{1-q}} \frac{H_-^2}{M_p^2} ~,~ \nu \equiv \frac{1-3q}{2(1-q)} ~,
\end{equation}
which is strongly blue. Correspondingly, we define the tensor-to-scalar ratio of the scalar field $\phi$, $r_\phi \equiv P_T/P_{\phi}$, therefore, at $\tau = \tau_-$ it is just a number $r_\phi = 96$. 

The dynamics in the bouncing phase is more involved. At first glance, the parameter $z_s$ diverges at the bouncing point $H=0$ according to \eqref{eq:zsgeneral}. However, if we use \eqref{eq:bouncephi}, the expression for $z_s$ becomes
\begin{equation}
    \frac{z_s^2}{a^2} = \frac{M_p^2}{4H^2} \frac{(1-\beta_1)^2}{3\beta_2} \left( e^{- \frac{t^2}{T^2}} - e^{- \frac{2t^2}{T^2}} \right) ~,
\end{equation}
which converges at the limit $t \to 0$ since $H = \alpha^2 t$ \footnote{The regularity of $z_s^2$ explains the prefactor $\sqrt{2(\beta_1 - 1)/3\beta_2}$ in \eqref{eq:bouncephi}. }. Moreover, at leading order we have $z_s/a = \rm{const}$ and thus $z_s^{\prime \prime}/z_s = a^{\prime \prime}/a$. In this case, the scalar fluctuation shares the same dynamics as the tensor fluctuation and as a result, the tensor-to-scalar ratio $r_\phi$ is invariant in the bouncing phase. We can thus immediately write down the power spectrum for $\phi$ at the end of bouncing phase:
\begin{equation}
\label{eq:Pphi}
    P_{\phi}(k,\tau_+) = P_T(k,\tau_+)/r_\phi = \frac{1}{96} P_T(k,\tau_+) ~.
\end{equation}

Before closing this section, we comment on the difference between our result and that in \cite{Cai:2012va,Zhu:2021whu}. The non-singular bouncing models in \cite{Cai:2012va,Zhu:2021whu} are constructed with a cubic Galileon action $\gamma G(X) \Box \phi$, in which the denominator in the expression of $z_s^2$ \eqref{eq:zsgeneral} become $(H M_p - \gamma \dot{\phi} G^{\prime}(X)/2)^2$. In this case, the dominant term in the denominator of \eqref{eq:zsgeneral} during bouncing phase is $(\gamma \dot{\phi} G^{\prime})^2$ instead of $H^2M_p^2$ in our scenario. Without the inclusion of cubic Galileon term, we simply get a constant tensor-to-scalar ratio $r_\phi = 96$ in our model.

\subsection{Scalar perturbation from curvaton field  $\sigma$}
As we show above, the single-field Ekpyrotic bouncing scenario predicts a blue-tilted scalar spectra, so we introduce the curvaton field $\sigma$ to get the scale-invariant power spectrum on CMB scale. The dynamical equation for the fluctuation of curvaton is
\begin{equation}
    u_k^{\prime \prime} + \left( k^2 - \frac{z_c^{\prime \prime}}{z_c} \right) u_k = 0 ~,
\end{equation}
where $u_k \equiv z_c \delta \sigma$ is the mode function for the curvaton field and 
\begin{equation}
    z_c \equiv a M_p f \sqrt{\frac{1 + \phi e^{\sqrt{\frac{2}{q}}\phi}}{e^{-\sqrt{\frac{2}{q}}\phi} + \phi^2 e^{\sqrt{\frac{2}{q}}\phi}}} \simeq \frac{aM_p f}{\sqrt{e^{-\sqrt{2/q} \phi}}} = aM_pf e^{\frac{\phi}{\sqrt{2q}}} ~,~ \phi \ll -1 ~.
\end{equation}
With the help of \eqref{eq:bgE} and \eqref{eq:ttau}, we have
\begin{equation}
    e^{\frac{\phi}{\sqrt{2q}}} = \frac{\sqrt{q(1-3q)}}{\sqrt{2V_0} M_p t} = - \frac{\sqrt{q(1-3q)}}{\sqrt{2V_0} M_p a_- (1-q)} \left[ \frac{(\tau_e-\tau)}{ (\tau_e - \tau_-)^q } \right]^{-\frac{1}{1-q}} ~.
\end{equation}
The background geometry \eqref{eq:bgcontraction} then implies
\begin{equation}
    z_c = \frac{-f\sqrt{q(1-3q)}}{\sqrt{2V_0} (1-q)} (\tau_e - \tau)^{-1} ~,
\end{equation}
so we get the following dynamical equation
\begin{equation}
\label{eq:modecurvaton}
    u_k^{\prime \prime} + \left( k^2 - \frac{2}{(\tau_e - \tau)^2} \right) u_k = 0 ~.
\end{equation}

Imposing the vacuum initial condition, the solution to \eqref{eq:modecurvaton} is
\begin{equation}
    u_k(\tau) = \frac{\sqrt{\pi (\tau_e - \tau)}}{2} H_{3/2}^{(1)} [k(\tau_e - \tau)] = \frac{e^{ik(\tau_e - \tau)}}{\sqrt{2k}} \left[ 1 + \frac{i}{k(\tau_e - \tau)} \right] ~,
\end{equation}
where $H_{3/2}^{(1)}$ is the Hankel function of the first kind. For modes that become super-horizon in the contraction epoch, the corresponding power spectrum at the end of contraction phase is
\begin{equation}
\label{eq:Psigmacontraction}
    P_{\sigma}(k,\tau_-) = \frac{k^3}{2\pi^2} \frac{|u_k|^2}{z_c^2} \simeq \frac{V_0(1-q)^2}{2\pi^2 f^2 q(1-3q) a_-^2} ~.
\end{equation}

In the bouncing epoch, we have
\begin{equation}
    \frac{z_c^{\prime \prime}}{z_c} \simeq \frac{a^{\prime \prime}}{a} - \frac{2}{q} \phi^{\prime 2} = \alpha^2 (1+\alpha^2 \tau^2) - \frac{4(\beta_1 - 1)}{3\beta_2 q} M_p^2 e^{-\frac{2\tau^2}{T^2} - \frac{1}{2} \alpha^2 \tau^2} ~,
\end{equation}
where we've applied \eqref{eq:bouncephi}. The dynamical equation near $\tau = 0$ becomes
\begin{equation}
    u_k^{\prime \prime} + \left[ k^2 - \tilde{\alpha}^2 + \tilde{\beta}^4 t^2 + \mathcal{O}(t^4) \right] u_k = 0 ~,~ 
\end{equation}
\begin{equation}
    \tilde{\alpha}^2 \equiv \alpha^2 - \frac{4(\beta_1 - 1)}{3\beta_2 q} M_p^2 ~,~ \tilde{\beta}^4 \equiv \alpha^4 + \frac{4(\beta_1 - 1)}{3\beta_2 q} M_p^2 \left( \frac{2}{T^2} + \frac{\alpha^2}{2} \right) ~.
\end{equation}
Expanding the Friedmann's equation \eqref{eq:Friedmann1} near the bouncing point gives
\begin{equation}
    3\alpha^4 t^2 M_p^2 \simeq \frac{(\beta_1 - 1)^2}{3\beta_2 T^2} M_p^2 t^2 ~\to~ \alpha^2 \simeq \frac{\beta_1 - 1}{3\sqrt{\beta_2}T} M_p ~.
\end{equation}
It's reasonable to take the duration of bouncing phase to be much larger than a Planck time in a classical bouncing scenario. As a result $T^{-1} < M_p$ and we have $\tilde{\alpha}^2 < 0$ (unless $\beta_2$ takes an extremely small value). Therefore, the curvaton acquires an oscillatory behavior in the bouncing phase, thus its power spectrum at $\tau_+$ is
\begin{equation}
\label{eq:Psigma}
    P_{\sigma}(k,\tau_+) = P_{\sigma}(k,\tau_-) = \frac{V_0(1-q)^2}{2\pi^2 f^2 q(1-3q) a_+^2} ~.
\end{equation}

\subsection{The curvature fluctuation}
\label{sec:r}
The gauge-invariant curvature fluctuation usually depends on the metric perturbation and the perturbed field $\delta \phi$ and $\delta \sigma$, although the former can be ignored by taking a spatial-flat gauge. In principle, one has to work out the explicit formalism for both the field perturbations to determine $\zeta$, which is rather involving. However, in curvaton scenarios we actually don't need to do that, since we effectively have only one field perturbation. In our case, however, which will be dominant depends on the scales we're considering.

From the above sections we can see that, $\delta\phi$ got a blue-tilted power spectrum while $\delta\sigma$ got a scale-invariant one, so it's natural that on small scales such as PTA scale, the dominate contribution from $P_{\zeta}$ is determined by the fluctuation of scalar field $\delta \phi$, whereas on large scales such as CMB scale, the curvature fluctuation $P_{\zeta}$ is mainly sourced from the curvaton field $\delta\sigma$. 

On small scales, since both background and perturbations comes from $\phi$ field, the system effectively becomes that of a single field, and the perturbations become adiabatic. From the above sections we know that the tensor-to-scalar ratio of $\phi$ is $r_{\phi} \equiv P_{\phi}/P_{T} = 96$. Moreover, as demonstrated above, the tensor power spectrum at the PTA frequency range ($f = 10{\rm nHz}$) is given by $P_T (f = 10{\rm nHz}) \sim 10^{-3}$,
which indicates the power spectrum of $P_{\phi}$ to be 
\begin{equation}
\label{eq:Pphiparametrize}
    P_{\phi}(k) = \frac{P_T}{96} \simeq 1.0 \times 10^{-5} \left( \frac{k}{10^6 {\rm Mpc}^{-1}} \right)^{\frac{2}{1-q}} ~.
\end{equation}
where we take the spectra index $n_T \equiv d\ln P_{T}/d\ln k$ to be approximately $2/(1-q)$ for illustrative purpose. 

Since on small scales the perturbations becomes adiabatic, the curvature pertuurbation generated from it is simply $\zeta \simeq -(H/\dot{\phi}) \delta \phi$. In the Ekpyrosis phase one has $H/\dot{\phi} = \sqrt{q/2}$, so
\begin{equation}
    P^{(s)}_\zeta(k, \tau_-)=\frac{q}{2}P_\phi(k,\tau_-)~,
\end{equation}
where the supscript ``$(s)$" denotes ``small scales". While since the bounce process is symmetric, it is reasonable to assume that $H/\dot{\phi}$ doesn't change much after the bounce, namely $P^{(s)}_\zeta(k, \tau_-)\simeq P^{(s)}_\zeta(k, \tau_+)$. Moreover, in the radiation dominated epoch, both the curvature and tensor fluctuations get frozen on super-horizon scales, so 
\begin{equation}
    \label{Pzeta-s}
    P^{(s)}_\zeta(k)\simeq P^{(s)}_\zeta(k, \tau_+)\sim 10^{-5} q \left( \frac{k}{10^6 {\rm Mpc}^{-1}} \right)^{\frac{2}{1-q}}~,
\end{equation}
where we've neglected numerical factors.

On large scales, on the other hand, the field fluctuation is solely determined by $P_{\sigma}$. However, the background is dominated by the $\phi$ field, instead of $\sigma$ field. Following the conventional treatment of curvaton mechanism, we assume that the curvaton field is converted into curvature fluctuation on its horizon re-entry event in the radiation dominated epoch, such that
\begin{equation}
    \zeta_k (t_r(k)) = -\frac{H \delta \sigma_k}{\dot{\sigma}} (t=t_r(k)) ~,
\end{equation}
where $t_r(k)$ labels the time of horizon re-entry event for a certain mode $k$. Since $t_r$ generically has $k$ dependence, a scale-invariant fluctuation of curvaton field doesn't necessariliy lead to a a scale-invariant curvature fluctuation. To generate a nearly scale-invariant curvature fluctuation on CMB scales, we are suggested to construct the model such that $H/\dot{\sigma} = {\rm const}$ in the radiation dominated epoch. 

The action of curvaton field \eqref{eq:Lsigma} has the schematic form $\mathcal{L}_{\sigma} = \mathcal{F}(\phi) (\partial \sigma)^2$, which gives the following equation of motion 
\begin{equation}
    \frac{d}{dt} \left( a^3 \mathcal{F}(\phi) \dot{\sigma} \right) = 0 ~.
\end{equation}
Except for the trivial solution $\dot{\sigma} = 0$, the evolving branch of solution is $\dot{\sigma} \propto a^{-3} \mathcal{F}(\phi)^{-1}$, and the condition becomes
\begin{equation}
\label{eq:curvatonconstraint}
    \frac{H}{\dot{\sigma}} = {\rm const} ~\to~ Ha^3 \mathcal{F}(\phi) = {\rm const} ~.
\end{equation}
In a radiation dominated universe, $a \propto t^{1/2}$, $H \propto t^{-1}$ and $\dot{\phi} \propto \sqrt{H} = t^{-1/2}$ from the Friedmann's equation $H^2 \propto X$. We then have $\phi \propto t^{1/2}$ and taking $\mathcal{F}(\phi) \propto \phi^{-1}$ could meet the condition \eqref{eq:curvatonconstraint}. This justifies the choice of curvation action \eqref{eq:Lsigma}. 

Since now $H/\dot\sigma$ is a constant with no time dependence, we define $s\equiv H/\dot\sigma|_{\tau=\tau_+}$ which is determined by the initial values of $H$ and $\dot\sigma$ in the radiation dominated epoch. This gives
\begin{equation}
    P^{(l)}_{\zeta} (k) = s^2 P_{\sigma} = \frac{s^2 V_0(1-q)^2}{2\pi^2 f^2 q(1-3q) a_+^2} ~,
\end{equation}
where the supscript ``$(l)$" denotes ``large scales". Moreover, when we compare this result to the CMB constraint, we have
\begin{equation}
    \label{Pzeta-l}
    P^{(l)}_{\zeta} (k_{\rm CMB}) = \frac{s^2 V_0(1-q)^2}{2\pi^2 f^2 q(1-3q) a_+^2} \simeq 2.1 \times 10^{-9} ~.
\end{equation}
and is scale-invariant till the re-entry of the horizon.

\begin{figure}[htp]
    \centering
    \includegraphics[width=0.8\linewidth]{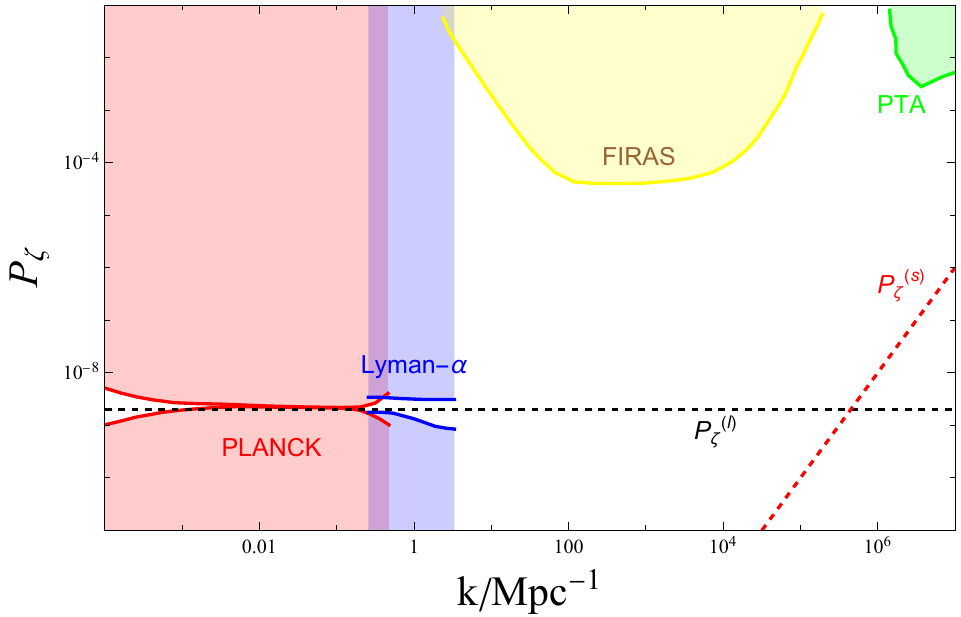}
    \caption{The power spectrum of $P^{s}_\zeta$ and $P^{l}_\zeta$, denoted by the red and black dashed lines respectively. The shaded region represents the constraint for the power spectrum of curvature perturbation $P_{\zeta}$ from Planck \cite{Planck:2018jri}, Lyman-$\alpha$ \cite{Bird:2010mp}, FIRAS \cite{Fixsen:1996nj, Chluba:2012we} (see also \cite{Chluba:2012gq,Cyr:2023pgw}) and PTA \cite{Byrnes:2018txb}. Here we take $q = 0.001$.}
    \label{fig:scalar}
\end{figure}
We depict the power spectrum of $P^{(s)}_\zeta$, $P^{(l)}_\zeta$ in Fig. \ref{fig:scalar}. It's easy to see that on the CMB scale, the curvature fluctuation $P_{\zeta}$ is mainly sourced from the curvaton field, whereas on the PTA range, the dominate contribution from $P_{\zeta}$ is determined by the fluctuation of scalar field $\delta \phi$. This is due to the fact that in Ekpyrotic models, the background will generate blue-tilted perturbations the same as the tensor perturbations, while the amplitude will be severely suppressed. Meanwhile, the curvaton field, which conformally couples to the background field in certain manner (see examples in \cite{Qiu:2011cy,Qiu:2013eoa}), will ``feel itself in an inflationary background", and generate the scale-invariant power spectrum as inflaton does. This is exactly the ``conformal mechanism" \cite{Hinterbichler:2011qk,Creminelli:2011mw} for generating spectrum required by the observational data. Moreover, taking $q=0.05$ as is allowed by PTA data, one can obtain the pivot scale where the perturbation of $\phi$ exceeds that of $\sigma$ by equaling Eqs. \eqref{Pzeta-s} and \eqref{Pzeta-l}. This will give rise to the result:
\begin{equation}
    k_{\rm pivot}\simeq 10^{-5}{\rm Mpc}^{-1}~.
\end{equation}

\section{Conclusion and Outlook}
\label{sec:Conclusion} 

The recently released PTA data of GW indicates that, if the GW is to be interpreted as from primordial tensor perturbation, then the tensor spectrum should be strongly blue-tilted. On the other hand, it is well-known that the Ekpyrosis-bounce scenario can provide tensor spectrum with spectral index $2<n_T<3$. So there is large possibility that such tensor perturbation may originate from the Ekpyrosis-bounce cosmology.

To illustrate this, in this paper we improve the study in \cite{Zhu:2023lbf} and present a concrete realization. Since the contracting phase preceding the bounce is Ekpyrotic-like where the EoS is much larger than unity, the anisotropic problem can be naturally removed. Moreover, although not much discussed in the current work, it is not difficult to understand that the ghost problem can also be alliminate by taking into account certain corrections, e.g., from Galileon theories. Bounce process are realized with numerical calculations, and finally, after the bounce, the universe safely evolves into a radiation-dominant phase, connecting with standard Big Bang cosmology.

The calculation of tensor perturbations is somehow straightforward. The tensor perturbations exit the horizon during the ekpyrotic phase, so their spectrum is blue-tilted. This feature can be inherited in radiation dominate phase, till the re-entry of the horizon in expanding phase. We have shown that after appropriately choosing model parameters, one can fit the PTA data very well. Moreover, we consider the constraints of the parameters against various constraints. The model is free of Trans-Planckian problem by requiring $H_+<M_p$. Moreover, considering the perturbation theory not being ruined, the bounce scenario can account for primordial GWs with wave number up to $10^7 {\rm Mpc}^{-1}$, containing the range of PTA detection. 

However, it is well known that the Ekpyrotic scenario also gives rise to blue spectrum to scalar perturbation, which conflicts with the CMB observations. In order to cure this problem, it is helpful to introduce a curvaton field $\sigma$, which couples to the Ekpyrotic field $\phi$. Due to the nontrivial coupling, the curvaton field can generate scale-invariant scalar perturbations, which will transfer to curvature perturbations after curvaton decay. We delicately design the form of coupling which can realize scale-invariance, and we also checked that the backreaction will not ruin the Ekpyrotic-bouncing background. Moreover, the blue-tilted perturbation of Ekpyrotic field will be suppressed on CMB scales, while be dominant on PTA scales. However, since it is still small compared to tensor perturbations with tiny tensor-to-scalar ratio, the problem of an oversized scalar fluctuation appeared in \cite{Zhu:2023lbf} is also solved in our improved model. 

A couple of future works are in order. First of all, we see that the scalar perturbation becomes dominant in small scales, which might have contributions on the matter power spectrum, and affect the consequent late-time evolution of perturbations and galaxy/large scale structure formation. Moreover, it seems that the current scalar perturbations are still not sufficient to generate Primordial Black Holes (PBHs), therefore it is also interesting to investigate how the PBHs can be generated from our scenario (see e.g., \cite{Chen:2022usd,Papanikolaou:2024fzf} for studies of PBH formation in bouncing cosmology, and see \cite{Sharma:2024whg} for other recent considerations). We will postpone this studies in a upcoming paper.

\acknowledgments
We thank Yan-chen Bi, Qing-guo Huang, Chunshan Lin, Guan-wen Yuan for useful discussions. T.Q. acknowledges Institute of Theoretical Physics, Chinese Academy of Sciences for its hospitality during his visit there. T.Q. is supported by the National Key Research and Development Program of China (Grant No. 2021YFC2203100), as well as Project 12047503 supported by National Natural Science Foundation of China. M.Z. was supported by grant No. UMO 2021/42/E/ST9/00260 from the National Science Centre, Poland.




\bibliography{bounce}

\providecommand{\href}[2]{#2}\begingroup\raggedright\begin{thebibliography}{100}

\bibitem{Novello:2008ra}
M.~Novello and S.~E.~P. Bergliaffa, ``{Bouncing Cosmologies},''
  \href{https://dx.doi.org/10.1016/j.physrep.2008.04.006}{{\em Phys. Rept.}
  {\bfseries 463} (2008) 127--213},
  \href{https://arxiv.org/abs/0802.1634}{{\ttfamily arXiv:0802.1634
  [astro-ph]}}.

\bibitem{Borde:1993xh}
A.~Borde and A.~Vilenkin, ``{Eternal inflation and the initial singularity},''
  \href{https://dx.doi.org/10.1103/PhysRevLett.72.3305}{{\em Phys. Rev. Lett.}
  {\bfseries 72} (1994) 3305--3309},
  \href{https://arxiv.org/abs/gr-qc/9312022}{{\ttfamily arXiv:gr-qc/9312022}}.

\bibitem{Borde:2001nh}
A.~Borde, A.~H. Guth, and A.~Vilenkin, ``{Inflationary space-times are
  incompletein past directions},''
  \href{https://dx.doi.org/10.1103/PhysRevLett.90.151301}{{\em Phys. Rev.
  Lett.} {\bfseries 90} (2003) 151301},
  \href{https://arxiv.org/abs/gr-qc/0110012}{{\ttfamily arXiv:gr-qc/0110012}}.

\bibitem{Lesnefsky:2022fen}
J.~E. Lesnefsky, D.~A. Easson, and P.~C.~W. Davies, ``{Past-completeness of
  inflationary spacetimes},''
  \href{https://dx.doi.org/10.1103/PhysRevD.107.044024}{{\em Phys. Rev. D}
  {\bfseries 107} no.~4, (2023) 044024},
  \href{https://arxiv.org/abs/2207.00955}{{\ttfamily arXiv:2207.00955
  [gr-qc]}}.

\bibitem{Geshnizjani:2023edw}
G.~Geshnizjani, E.~Ling, and J.~Quintin, ``{On the initial singularity and
  extendibility of flat quasi-de Sitter spacetimes},''
  \href{https://arxiv.org/abs/2305.01676}{{\ttfamily arXiv:2305.01676
  [gr-qc]}}.

\bibitem{Brandenberger:2000wr}
R.~H. Brandenberger and J.~Martin, ``{The Robustness of inflation to changes in
  superPlanck scale physics},''
  \href{https://dx.doi.org/10.1142/S0217732301004170}{{\em Mod. Phys. Lett. A}
  {\bfseries 16} (2001) 999--1006},
  \href{https://arxiv.org/abs/astro-ph/0005432}{{\ttfamily
  arXiv:astro-ph/0005432}}.

\bibitem{Martin:2000xs}
J.~Martin and R.~H. Brandenberger, ``{The TransPlanckian problem of
  inflationary cosmology},''
  \href{https://dx.doi.org/10.1103/PhysRevD.63.123501}{{\em Phys. Rev. D}
  {\bfseries 63} (2001) 123501},
  \href{https://arxiv.org/abs/hep-th/0005209}{{\ttfamily
  arXiv:hep-th/0005209}}.

\bibitem{Ijjas:2018qbo}
A.~Ijjas and P.~J. Steinhardt, ``{Bouncing Cosmology made simple},''
  \href{https://dx.doi.org/10.1088/1361-6382/aac482}{{\em Class. Quant. Grav.}
  {\bfseries 35} no.~13, (2018) 135004},
  \href{https://arxiv.org/abs/1803.01961}{{\ttfamily arXiv:1803.01961
  [astro-ph.CO]}}.

\bibitem{NANOGrav:2023hvm}
{\bfseries NANOGrav} Collaboration, A.~Afzal {\em et~al.}, ``{The NANOGrav 15
  yr Data Set: Search for Signals from New Physics},''
  \href{https://dx.doi.org/10.3847/2041-8213/acdc91}{{\em Astrophys. J. Lett.}
  {\bfseries 951} no.~1, (2023) L11},
  \href{https://arxiv.org/abs/2306.16219}{{\ttfamily arXiv:2306.16219
  [astro-ph.HE]}}.

\bibitem{NANOGrav:2023gor}
{\bfseries NANOGrav} Collaboration, G.~Agazie {\em et~al.}, ``{The NANOGrav 15
  yr Data Set: Evidence for a Gravitational-wave Background},''
  \href{https://dx.doi.org/10.3847/2041-8213/acdac6}{{\em Astrophys. J. Lett.}
  {\bfseries 951} no.~1, (2023) L8},
  \href{https://arxiv.org/abs/2306.16213}{{\ttfamily arXiv:2306.16213
  [astro-ph.HE]}}.

\bibitem{Antoniadis:2023rey}
J.~Antoniadis {\em et~al.}, ``{The second data release from the European Pulsar
  Timing Array III. Search for gravitational wave signals},''
  \href{https://arxiv.org/abs/2306.16214}{{\ttfamily arXiv:2306.16214
  [astro-ph.HE]}}.

\bibitem{Reardon:2023gzh}
D.~J. Reardon {\em et~al.}, ``{Search for an Isotropic Gravitational-wave
  Background with the Parkes Pulsar Timing Array},''
  \href{https://dx.doi.org/10.3847/2041-8213/acdd02}{{\em Astrophys. J. Lett.}
  {\bfseries 951} no.~1, (2023) L6},
  \href{https://arxiv.org/abs/2306.16215}{{\ttfamily arXiv:2306.16215
  [astro-ph.HE]}}.

\bibitem{Xu:2023wog}
H.~Xu {\em et~al.}, ``{Searching for the Nano-Hertz Stochastic Gravitational
  Wave Background with the Chinese Pulsar Timing Array Data Release I},''
  \href{https://dx.doi.org/10.1088/1674-4527/acdfa5}{{\em Res. Astron.
  Astrophys.} {\bfseries 23} no.~7, (2023) 075024},
  \href{https://arxiv.org/abs/2306.16216}{{\ttfamily arXiv:2306.16216
  [astro-ph.HE]}}.

\bibitem{Battista:2021rlh}
E.~Battista and V.~De~Falco, ``{First post-Newtonian generation of
  gravitational waves in Einstein-Cartan theory},''
  \href{https://dx.doi.org/10.1103/PhysRevD.104.084067}{{\em Phys. Rev. D}
  {\bfseries 104} no.~8, (2021) 084067},
  \href{https://arxiv.org/abs/2109.01384}{{\ttfamily arXiv:2109.01384
  [gr-qc]}}.

\bibitem{Battista:2022hmv}
E.~Battista and V.~De~Falco, ``{Gravitational waves at the first post-Newtonian
  order with the Weyssenhoff fluid in Einstein\textendash{}Cartan theory},''
  \href{https://dx.doi.org/10.1140/epjc/s10052-022-10558-9}{{\em Eur. Phys. J.
  C} {\bfseries 82} no.~7, (2022) 628},
  \href{https://arxiv.org/abs/2206.12907}{{\ttfamily arXiv:2206.12907
  [gr-qc]}}.

\bibitem{Ashoorioon:2022raz}
A.~Ashoorioon, K.~Rezazadeh, and A.~Rostami, ``{NANOGrav signal from the end of
  inflation and the LIGO mass and heavier primordial black holes},''
  \href{https://dx.doi.org/10.1016/j.physletb.2022.137542}{{\em Phys. Lett. B}
  {\bfseries 835} (2022) 137542},
  \href{https://arxiv.org/abs/2202.01131}{{\ttfamily arXiv:2202.01131
  [astro-ph.CO]}}.

\bibitem{Ellis:2023tsl}
J.~Ellis, M.~Lewicki, C.~Lin, and V.~Vaskonen, ``{Cosmic superstrings revisited
  in light of NANOGrav 15-year data},''
  \href{https://dx.doi.org/10.1103/PhysRevD.108.103511}{{\em Phys. Rev. D}
  {\bfseries 108} no.~10, (2023) 103511},
  \href{https://arxiv.org/abs/2306.17147}{{\ttfamily arXiv:2306.17147
  [astro-ph.CO]}}.

\bibitem{Ellis:2023dgf}
J.~Ellis, M.~Fairbairn, G.~H\"utsi, J.~Raidal, J.~Urrutia, V.~Vaskonen, and
  H.~Veerm\"ae, ``{Gravitational waves from supermassive black hole binaries in
  light of the NANOGrav 15-year data},''
  \href{https://dx.doi.org/10.1103/PhysRevD.109.L021302}{{\em Phys. Rev. D}
  {\bfseries 109} no.~2, (2024) L021302},
  \href{https://arxiv.org/abs/2306.17021}{{\ttfamily arXiv:2306.17021
  [astro-ph.CO]}}.

\bibitem{Franciolini:2023pbf}
G.~Franciolini, A.~Iovino, Junior., V.~Vaskonen, and H.~Veermae, ``{Recent
  Gravitational Wave Observation by Pulsar Timing Arrays and Primordial Black
  Holes: The Importance of Non-Gaussianities},''
  \href{https://dx.doi.org/10.1103/PhysRevLett.131.201401}{{\em Phys. Rev.
  Lett.} {\bfseries 131} no.~20, (2023) 201401},
  \href{https://arxiv.org/abs/2306.17149}{{\ttfamily arXiv:2306.17149
  [astro-ph.CO]}}.

\bibitem{Zhang:2023lzt}
C.~Zhang, N.~Dai, Q.~Gao, Y.~Gong, T.~Jiang, and X.~Lu, ``{Detecting new
  fundamental fields with pulsar timing arrays},''
  \href{https://dx.doi.org/10.1103/PhysRevD.108.104069}{{\em Phys. Rev. D}
  {\bfseries 108} no.~10, (2023) 104069},
  \href{https://arxiv.org/abs/2307.01093}{{\ttfamily arXiv:2307.01093
  [gr-qc]}}.

\bibitem{Cannizzaro:2023mgc}
E.~Cannizzaro, G.~Franciolini, and P.~Pani, ``{Novel tests of gravity using
  nano-Hertz stochastic gravitational-wave background signals},''
  \href{https://dx.doi.org/10.1088/1475-7516/2024/04/056}{{\em JCAP} {\bfseries
  04} (2024) 056}, \href{https://arxiv.org/abs/2307.11665}{{\ttfamily
  arXiv:2307.11665 [gr-qc]}}.

\bibitem{Shen:2023pan}
Z.-Q. Shen, G.-W. Yuan, Y.-Y. Wang, and Y.-Z. Wang, ``{Dark Matter Spike
  surrounding Supermassive Black Holes Binary and the nanohertz Stochastic
  Gravitational Wave Background},''
  \href{https://arxiv.org/abs/2306.17143}{{\ttfamily arXiv:2306.17143
  [astro-ph.HE]}}.

\bibitem{Du:2023qvj}
X.~K. Du, M.~X. Huang, F.~Wang, and Y.~K. Zhang, ``{Did the nHZ Gravitational
  Waves Signatures Observed By NANOGrav Indicate Multiple Sector SUSY
  Breaking?},'' \href{https://arxiv.org/abs/2307.02938}{{\ttfamily
  arXiv:2307.02938 [hep-ph]}}.

\bibitem{Balaji:2023ehk}
S.~Balaji, G.~Dom\`enech, and G.~Franciolini, ``{Scalar-induced gravitational
  wave interpretation of PTA data: the role of scalar fluctuation propagation
  speed},'' \href{https://dx.doi.org/10.1088/1475-7516/2023/10/041}{{\em JCAP}
  {\bfseries 10} (2023) 041},
  \href{https://arxiv.org/abs/2307.08552}{{\ttfamily arXiv:2307.08552
  [gr-qc]}}.

\bibitem{Zhang:2023nrs}
Z.~Zhang, C.~Cai, Y.-H. Su, S.~Wang, Z.-H. Yu, and H.-H. Zhang, ``{Nano-Hertz
  gravitational waves from collapsing domain walls associated with freeze-in
  dark matter in light of pulsar timing array observations},''
  \href{https://dx.doi.org/10.1103/PhysRevD.108.095037}{{\em Phys. Rev. D}
  {\bfseries 108} no.~9, (2023) 095037},
  \href{https://arxiv.org/abs/2307.11495}{{\ttfamily arXiv:2307.11495
  [hep-ph]}}.

\bibitem{Konoplya:2023fmh}
R.~A. Konoplya and A.~Zhidenko, ``{Asymptotic tails of massive gravitons in
  light of pulsar timing array observations},''
  \href{https://dx.doi.org/10.1016/j.physletb.2024.138685}{{\em Phys. Lett. B}
  {\bfseries 853} (2024) 138685},
  \href{https://arxiv.org/abs/2307.01110}{{\ttfamily arXiv:2307.01110
  [gr-qc]}}.

\bibitem{Wu:2023hsa}
Y.-M. Wu, Z.-C. Chen, and Q.-G. Huang, ``{Cosmological interpretation for the
  stochastic signal in pulsar timing arrays},''
  \href{https://dx.doi.org/10.1007/s11433-023-2298-7}{{\em Sci. China Phys.
  Mech. Astron.} {\bfseries 67} no.~4, (2024) 240412},
  \href{https://arxiv.org/abs/2307.03141}{{\ttfamily arXiv:2307.03141
  [astro-ph.CO]}}.

\bibitem{Ghosh:2023aum}
T.~Ghosh, A.~Ghoshal, H.-K. Guo, F.~Hajkarim, S.~F. King, K.~Sinha, X.~Wang,
  and G.~White, ``{Did we hear the sound of the Universe boiling? Analysis
  using the full fluid velocity profiles and NANOGrav 15-year data},''
  \href{https://dx.doi.org/10.1088/1475-7516/2024/05/100}{{\em JCAP} {\bfseries
  05} (2024) 100}, \href{https://arxiv.org/abs/2307.02259}{{\ttfamily
  arXiv:2307.02259 [astro-ph.HE]}}.

\bibitem{Huang:2023chx}
H.-L. Huang, Y.~Cai, J.-Q. Jiang, J.~Zhang, and Y.-S. Piao, ``{Supermassive
  primordial black holes in multiverse: for nano-Hertz gravitational wave and
  high-redshift JWST galaxies},''
  \href{https://arxiv.org/abs/2306.17577}{{\ttfamily arXiv:2306.17577
  [gr-qc]}}.

\bibitem{Liu:2023pau}
L.~Liu, Z.-C. Chen, and Q.-G. Huang, ``{Probing the equation of state of the
  early Universe with pulsar timing arrays},''
  \href{https://dx.doi.org/10.1088/1475-7516/2023/11/071}{{\em JCAP} {\bfseries
  11} (2023) 071}, \href{https://arxiv.org/abs/2307.14911}{{\ttfamily
  arXiv:2307.14911 [astro-ph.CO]}}.

\bibitem{Cai:2023dls}
Y.-F. Cai, X.-C. He, X.-H. Ma, S.-F. Yan, and G.-W. Yuan, ``{Limits on
  scalar-induced gravitational waves from the stochastic background by pulsar
  timing array observations},''
  \href{https://dx.doi.org/10.1016/j.scib.2023.10.027}{{\em Sci. Bull.}
  {\bfseries 68} (2023) 2929--2935},
  \href{https://arxiv.org/abs/2306.17822}{{\ttfamily arXiv:2306.17822
  [gr-qc]}}.

\bibitem{Liu:2023ymk}
L.~Liu, Z.-C. Chen, and Q.-G. Huang, ``{Implications for the non-Gaussianity of
  curvature perturbation from pulsar timing arrays},''
  \href{https://dx.doi.org/10.1103/PhysRevD.109.L061301}{{\em Phys. Rev. D}
  {\bfseries 109} no.~6, (2024) L061301},
  \href{https://arxiv.org/abs/2307.01102}{{\ttfamily arXiv:2307.01102
  [astro-ph.CO]}}.

\bibitem{Jiang:2023qbm}
S.~Jiang, A.~Yang, J.~Ma, and F.~P. Huang, ``{Implication of nano-Hertz
  stochastic gravitational wave on dynamical dark matter through a dark
  first-order phase transition},''
  \href{https://dx.doi.org/10.1088/1361-6382/ad24c6}{{\em Class. Quant. Grav.}
  {\bfseries 41} no.~6, (2024) 065009},
  \href{https://arxiv.org/abs/2306.17827}{{\ttfamily arXiv:2306.17827
  [hep-ph]}}.

\bibitem{Cai:2023uhc}
Y.~Cai, M.~Zhu, and Y.-S. Piao, ``{Primordial Black Holes from Null Energy
  Condition Violation during Inflation},''
  \href{https://dx.doi.org/10.1103/PhysRevLett.133.021001}{{\em Phys. Rev.
  Lett.} {\bfseries 133} no.~2, (2024) 021001},
  \href{https://arxiv.org/abs/2305.10933}{{\ttfamily arXiv:2305.10933
  [gr-qc]}}.

\bibitem{Maji:2023fhv}
R.~Maji and W.-I. Park, ``{Supersymmetric U(1)B-L flat direction and NANOGrav
  15 year data},'' \href{https://dx.doi.org/10.1088/1475-7516/2024/01/015}{{\em
  JCAP} {\bfseries 01} (2024) 015},
  \href{https://arxiv.org/abs/2308.11439}{{\ttfamily arXiv:2308.11439
  [hep-ph]}}.

\bibitem{Addazi:2023jvg}
A.~Addazi, Y.-F. Cai, A.~Marciano, and L.~Visinelli, ``{Have pulsar timing
  array methods detected a cosmological phase transition?},''
  \href{https://dx.doi.org/10.1103/PhysRevD.109.015028}{{\em Phys. Rev. D}
  {\bfseries 109} no.~1, (2024) 015028},
  \href{https://arxiv.org/abs/2306.17205}{{\ttfamily arXiv:2306.17205
  [astro-ph.CO]}}.

\bibitem{Ellis:2023oxs}
J.~Ellis, M.~Fairbairn, G.~Franciolini, G.~H\"utsi, A.~Iovino, M.~Lewicki,
  M.~Raidal, J.~Urrutia, V.~Vaskonen, and H.~Veerm\"ae, ``{What is the source
  of the PTA GW signal?},''
  \href{https://dx.doi.org/10.1103/PhysRevD.109.023522}{{\em Phys. Rev. D}
  {\bfseries 109} no.~2, (2024) 023522},
  \href{https://arxiv.org/abs/2308.08546}{{\ttfamily arXiv:2308.08546
  [astro-ph.CO]}}.

\bibitem{Li:2023bxy}
S.-P. Li and K.-P. Xie, ``{Collider test of nano-Hertz gravitational waves from
  pulsar timing arrays},''
  \href{https://dx.doi.org/10.1103/PhysRevD.108.055018}{{\em Phys. Rev. D}
  {\bfseries 108} no.~5, (2023) 055018},
  \href{https://arxiv.org/abs/2307.01086}{{\ttfamily arXiv:2307.01086
  [hep-ph]}}.

\bibitem{Xiao:2023dbb}
Y.~Xiao, J.~M. Yang, and Y.~Zhang, ``{Implications of nano-Hertz gravitational
  waves on electroweak phase transition in the singlet dark matter model},''
  \href{https://dx.doi.org/10.1016/j.scib.2023.11.025}{{\em Sci. Bull.}
  {\bfseries 68} (2023) 3158--3164},
  \href{https://arxiv.org/abs/2307.01072}{{\ttfamily arXiv:2307.01072
  [hep-ph]}}.

\bibitem{Han:2023olf}
C.~Han, K.-P. Xie, J.~M. Yang, and M.~Zhang, ``{Self-interacting dark matter
  implied by nano-Hertz gravitational waves},''
  \href{https://dx.doi.org/10.1103/PhysRevD.109.115025}{{\em Phys. Rev. D}
  {\bfseries 109} no.~11, (2024) 115025},
  \href{https://arxiv.org/abs/2306.16966}{{\ttfamily arXiv:2306.16966
  [hep-ph]}}.

\bibitem{Oikonomou:2023bah}
V.~K. Oikonomou, ``{Effects of the axion through the Higgs portal on primordial
  gravitational waves during the electroweak breaking},''
  \href{https://dx.doi.org/10.1103/PhysRevD.107.064071}{{\em Phys. Rev. D}
  {\bfseries 107} no.~6, (2023) 064071},
  \href{https://arxiv.org/abs/2303.05889}{{\ttfamily arXiv:2303.05889
  [hep-ph]}}.

\bibitem{Bian:2023dnv}
L.~Bian, S.~Ge, J.~Shu, B.~Wang, X.-Y. Yang, and J.~Zong, ``{Gravitational wave
  sources for pulsar timing arrays},''
  \href{https://dx.doi.org/10.1103/PhysRevD.109.L101301}{{\em Phys. Rev. D}
  {\bfseries 109} no.~10, (2024) L101301},
  \href{https://arxiv.org/abs/2307.02376}{{\ttfamily arXiv:2307.02376
  [astro-ph.HE]}}.

\bibitem{Huang:2023zvs}
M.~X. Huang, F.~Wang, and Y.~K. Zhang, ``{Interplay between the muon g-2
  anomaly and the PTA nHZ gravitational waves from domain walls in the
  next-to-minimal supersymmetric standard model},''
  \href{https://dx.doi.org/10.1103/PhysRevD.109.075032}{{\em Phys. Rev. D}
  {\bfseries 109} no.~7, (2024) 075032},
  \href{https://arxiv.org/abs/2309.06378}{{\ttfamily arXiv:2309.06378
  [hep-ph]}}.

\bibitem{Ye:2023xyr}
G.~Ye and A.~Silvestri, ``{Can the Gravitational Wave Background Feel Wiggles
  in Spacetime?},'' \href{https://dx.doi.org/10.3847/2041-8213/ad2851}{{\em
  Astrophys. J. Lett.} {\bfseries 963} no.~1, (2024) L15},
  \href{https://arxiv.org/abs/2307.05455}{{\ttfamily arXiv:2307.05455
  [astro-ph.CO]}}.

\bibitem{Guo:2023hyp}
S.-Y. Guo, M.~Khlopov, X.~Liu, L.~Wu, Y.~Wu, and B.~Zhu, ``{Footprints of
  Axion-Like Particle in Pulsar Timing Array Data and JWST Observations},''
  \href{https://arxiv.org/abs/2306.17022}{{\ttfamily arXiv:2306.17022
  [hep-ph]}}.

\bibitem{Choudhury:2023kam}
S.~Choudhury, ``{Single field inflation in the light of Pulsar Timing Array
  Data: quintessential interpretation of blue tilted tensor spectrum through
  Non-Bunch Davies initial condition},''
  \href{https://dx.doi.org/10.1140/epjc/s10052-024-12625-9}{{\em Eur. Phys. J.
  C} {\bfseries 84} no.~3, (2024) 278},
  \href{https://arxiv.org/abs/2307.03249}{{\ttfamily arXiv:2307.03249
  [astro-ph.CO]}}.

\bibitem{Jiang:2023gfe}
J.-Q. Jiang, Y.~Cai, G.~Ye, and Y.-S. Piao, ``{Broken blue-tilted inflationary
  gravitational waves: a joint analysis of NANOGrav 15-year and BICEP/Keck 2018
  data},'' \href{https://dx.doi.org/10.1088/1475-7516/2024/05/004}{{\em JCAP}
  {\bfseries 05} (2024) 004},
  \href{https://arxiv.org/abs/2307.15547}{{\ttfamily arXiv:2307.15547
  [astro-ph.CO]}}.

\bibitem{Oikonomou:2023bli}
V.~K. Oikonomou, ``{A Stiff Pre-CMB Era with a Mildly Blue-tilted Tensor
  Inflationary Era can Explain the 2023 NANOGrav Signal},''
  \href{https://arxiv.org/abs/2309.04850}{{\ttfamily arXiv:2309.04850
  [astro-ph.CO]}}.

\bibitem{Choudhury:2023fwk}
S.~Choudhury, K.~Dey, A.~Karde, S.~Panda, and M.~Sami, ``{Primordial
  non-Gaussianity as a saviour for PBH overproduction in SIGWs generated by
  pulsar timing arrays for Galileon inflation},''
  \href{https://dx.doi.org/10.1016/j.physletb.2024.138925}{{\em Phys. Lett. B}
  {\bfseries 856} (2024) 138925},
  \href{https://arxiv.org/abs/2310.11034}{{\ttfamily arXiv:2310.11034
  [astro-ph.CO]}}.

\bibitem{Lozanov:2023rcd}
K.~D. Lozanov, S.~Pi, M.~Sasaki, V.~Takhistov, and A.~Wang, ``{Axion Universal
  Gravitational Wave Interpretation of Pulsar Timing Array Data},''
  \href{https://arxiv.org/abs/2310.03594}{{\ttfamily arXiv:2310.03594
  [astro-ph.CO]}}.

\bibitem{Oikonomou:2023qfz}
V.~K. Oikonomou, ``{Flat energy spectrum of primordial gravitational waves
  versus peaks and the NANOGrav 2023 observation},''
  \href{https://dx.doi.org/10.1103/PhysRevD.108.043516}{{\em Phys. Rev. D}
  {\bfseries 108} no.~4, (2023) 043516},
  \href{https://arxiv.org/abs/2306.17351}{{\ttfamily arXiv:2306.17351
  [astro-ph.CO]}}.

\bibitem{Choudhury:2023fjs}
S.~Choudhury, K.~Dey, and A.~Karde, ``{Untangling PBH overproduction in
  $w$-SIGWs generated by Pulsar Timing Arrays for MST-EFT of single field
  inflation},'' \href{https://arxiv.org/abs/2311.15065}{{\ttfamily
  arXiv:2311.15065 [astro-ph.CO]}}.

\bibitem{Liu:2023hpw}
L.~Liu, Y.~Wu, and Z.-C. Chen, ``{Simultaneously probing the sound speed and
  equation of state of the early Universe with pulsar timing arrays},''
  \href{https://dx.doi.org/10.1088/1475-7516/2024/04/011}{{\em JCAP} {\bfseries
  04} (2024) 011}, \href{https://arxiv.org/abs/2310.16500}{{\ttfamily
  arXiv:2310.16500 [astro-ph.CO]}}.

\bibitem{Ye:2023tpz}
G.~Ye, M.~Zhu, and Y.~Cai, ``{Null energy condition violation during inflation
  and pulsar timing array observations},''
  \href{https://dx.doi.org/10.1007/JHEP02(2024)008}{{\em JHEP} {\bfseries 02}
  (2024) 008}, \href{https://arxiv.org/abs/2312.10685}{{\ttfamily
  arXiv:2312.10685 [gr-qc]}}.

\bibitem{Chowdhury:2023xvy}
D.~Chowdhury, A.~Hait, S.~Mohanty, and S.~Prakash, ``{Ultralight
  $(L_\mu-L_\tau)$ vector dark matter interpretation of NANOGrav
  observations},'' \href{https://arxiv.org/abs/2311.10148}{{\ttfamily
  arXiv:2311.10148 [hep-ph]}}.

\bibitem{Chen:2024fir}
Z.-C. Chen, J.~Li, L.~Liu, and Z.~Yi, ``{Probing the speed of scalar-induced
  gravitational waves with pulsar timing arrays},''
  \href{https://dx.doi.org/10.1103/PhysRevD.109.L101302}{{\em Phys. Rev. D}
  {\bfseries 109} no.~10, (2024) L101302},
  \href{https://arxiv.org/abs/2401.09818}{{\ttfamily arXiv:2401.09818
  [gr-qc]}}.

\bibitem{Jiang:2024dxj}
J.-Q. Jiang and Y.-S. Piao, ``{Search for the non-linearities of gravitational
  wave background in NANOGrav 15-year data set},''
  \href{https://arxiv.org/abs/2401.16950}{{\ttfamily arXiv:2401.16950
  [gr-qc]}}.

\bibitem{Chen:2024twp}
Z.-C. Chen and L.~Liu, ``{Can we distinguish the adiabatic fluctuations and
  isocurvature fluctuations with pulsar timing arrays?},''
  \href{https://arxiv.org/abs/2402.16781}{{\ttfamily arXiv:2402.16781
  [astro-ph.CO]}}.

\bibitem{Choudhury:2024dzw}
S.~Choudhury, S.~Ganguly, S.~Panda, S.~SenGupta, and P.~Tiwari, ``{Obviating
  PBH overproduction for SIGWs generated by Pulsar Timing Arrays in loop
  corrected EFT of bounce},''
  \href{https://arxiv.org/abs/2407.18976}{{\ttfamily arXiv:2407.18976
  [astro-ph.CO]}}.

\bibitem{Chen:2024mwg}
Z.-C. Chen and L.~Liu, ``{Constraints on inflation with null energy condition
  violation from advanced LIGO and advanced Virgo's first three observing
  runs},'' \href{https://dx.doi.org/10.1088/1475-7516/2024/06/028}{{\em JCAP}
  {\bfseries 06} (2024) 028},
  \href{https://arxiv.org/abs/2404.07075}{{\ttfamily arXiv:2404.07075
  [gr-qc]}}.

\bibitem{Li:2024dce}
C.~Li, ``{Primordial Gravitational Waves of Big Bounce Cosmology in Light of
  Stochastic Gravitational Wave Background},''
  \href{https://arxiv.org/abs/2407.10071}{{\ttfamily arXiv:2407.10071
  [astro-ph.CO]}}.

\bibitem{Chen:2024jca}
Z.-C. Chen and L.~Liu, ``{Detecting a Gravitational-Wave Background from Null
  Energy Condition Violation: Prospects for Taiji},''
  \href{https://arxiv.org/abs/2404.08375}{{\ttfamily arXiv:2404.08375
  [gr-qc]}}.

\bibitem{Pallis:2024mip}
C.~Pallis, ``{PeV-Scale SUSY and Cosmic Strings from F-Term Hybrid
  Inflation},'' \href{https://dx.doi.org/10.3390/universe10050211}{{\em
  Universe} {\bfseries 10} no.~5, (2024) },
  \href{https://arxiv.org/abs/2403.09385}{{\ttfamily arXiv:2403.09385
  [hep-ph]}}.

\bibitem{Li:2024oru}
C.~Li, J.~Lai, J.~Xiang, and C.~Wu, ``{Dual Inflation and Bounce Cosmologies
  Interpretation of Pulsar Timing Array Data},''
  \href{https://arxiv.org/abs/2405.15889}{{\ttfamily arXiv:2405.15889
  [astro-ph.CO]}}.

\bibitem{Vagnozzi:2023lwo}
S.~Vagnozzi, ``{Inflationary interpretation of the stochastic gravitational
  wave background signal detected by pulsar timing array experiments},''
  \href{https://dx.doi.org/10.1016/j.jheap.2023.07.001}{{\em JHEAp} {\bfseries
  39} (2023) 81--98}, \href{https://arxiv.org/abs/2306.16912}{{\ttfamily
  arXiv:2306.16912 [astro-ph.CO]}}.

\bibitem{Wang:2014kqa}
Y.~Wang and W.~Xue, ``{Inflation and Alternatives with Blue Tensor Spectra},''
  \href{https://dx.doi.org/10.1088/1475-7516/2014/10/075}{{\em JCAP} {\bfseries
  10} (2014) 075}, \href{https://arxiv.org/abs/1403.5817}{{\ttfamily
  arXiv:1403.5817 [astro-ph.CO]}}.

\bibitem{Khoury:2001wf}
J.~Khoury, B.~A. Ovrut, P.~J. Steinhardt, and N.~Turok, ``{The Ekpyrotic
  universe: Colliding branes and the origin of the hot big bang},''
  \href{https://dx.doi.org/10.1103/PhysRevD.64.123522}{{\em Phys. Rev. D}
  {\bfseries 64} (2001) 123522},
  \href{https://arxiv.org/abs/hep-th/0103239}{{\ttfamily
  arXiv:hep-th/0103239}}.

\bibitem{Brandenberger:2009jq}
R.~H. Brandenberger, ``{Alternatives to the inflationary paradigm of structure
  formation},'' \href{https://dx.doi.org/10.1142/S2010194511000109}{{\em Int.
  J. Mod. Phys. Conf. Ser.} {\bfseries 01} (2011) 67--79},
  \href{https://arxiv.org/abs/0902.4731}{{\ttfamily arXiv:0902.4731 [hep-th]}}.

\bibitem{Brandenberger:2020tcr}
R.~Brandenberger and Z.~Wang, ``{Nonsingular Ekpyrotic Cosmology with a Nearly
  Scale-Invariant Spectrum of Cosmological Perturbations and Gravitational
  Waves},'' \href{https://dx.doi.org/10.1103/PhysRevD.101.063522}{{\em Phys.
  Rev. D} {\bfseries 101} no.~6, (2020) 063522},
  \href{https://arxiv.org/abs/2001.00638}{{\ttfamily arXiv:2001.00638
  [hep-th]}}.

\bibitem{Qiu:2011cy}
T.~Qiu, J.~Evslin, Y.-F. Cai, M.~Li, and X.~Zhang, ``{Bouncing Galileon
  Cosmologies},'' \href{https://dx.doi.org/10.1088/1475-7516/2011/10/036}{{\em
  JCAP} {\bfseries 10} (2011) 036},
  \href{https://arxiv.org/abs/1108.0593}{{\ttfamily arXiv:1108.0593 [hep-th]}}.

\bibitem{Cai:2012ag}
Y.-F. Cai, C.~Gao, and E.~N. Saridakis, ``{Bounce and cyclic cosmology in
  extended nonlinear massive gravity},''
  \href{https://dx.doi.org/10.1088/1475-7516/2012/10/048}{{\em JCAP} {\bfseries
  10} (2012) 048}, \href{https://arxiv.org/abs/1207.3786}{{\ttfamily
  arXiv:1207.3786 [astro-ph.CO]}}.

\bibitem{Cai:2012va}
Y.-F. Cai, D.~A. Easson, and R.~Brandenberger, ``{Towards a Nonsingular
  Bouncing Cosmology},''
  \href{https://dx.doi.org/10.1088/1475-7516/2012/08/020}{{\em JCAP} {\bfseries
  08} (2012) 020}, \href{https://arxiv.org/abs/1206.2382}{{\ttfamily
  arXiv:1206.2382 [hep-th]}}.

\bibitem{Cai:2014jla}
Y.-F. Cai and E.~Wilson-Ewing, ``{A $\Lambda$CDM bounce scenario},''
  \href{https://dx.doi.org/10.1088/1475-7516/2015/03/006}{{\em JCAP} {\bfseries
  03} (2015) 006}, \href{https://arxiv.org/abs/1412.2914}{{\ttfamily
  arXiv:1412.2914 [gr-qc]}}.

\bibitem{Qiu:2015nha}
T.~Qiu and Y.-T. Wang, ``{G-Bounce Inflation: Towards Nonsingular Inflation
  Cosmology with Galileon Field},''
  \href{https://dx.doi.org/10.1007/JHEP04(2015)130}{{\em JHEP} {\bfseries 04}
  (2015) 130}, \href{https://arxiv.org/abs/1501.03568}{{\ttfamily
  arXiv:1501.03568 [astro-ph.CO]}}.

\bibitem{Wan:2015hya}
Y.~Wan, T.~Qiu, F.~P. Huang, Y.-F. Cai, H.~Li, and X.~Zhang, ``{Bounce
  Inflation Cosmology with Standard Model Higgs Boson},''
  \href{https://dx.doi.org/10.1088/1475-7516/2015/12/019}{{\em JCAP} {\bfseries
  12} (2015) 019}, \href{https://arxiv.org/abs/1509.08772}{{\ttfamily
  arXiv:1509.08772 [gr-qc]}}.

\bibitem{Li:2016awk}
H.-G. Li, Y.~Cai, and Y.-S. Piao, ``{Towards the bounce inflationary
  gravitational wave},''
  \href{https://dx.doi.org/10.1140/epjc/s10052-016-4554-2}{{\em Eur. Phys. J.
  C} {\bfseries 76} no.~12, (2016) 699},
  \href{https://arxiv.org/abs/1605.09586}{{\ttfamily arXiv:1605.09586
  [gr-qc]}}.

\bibitem{Cai:2016hea}
Y.-F. Cai, A.~Marciano, D.-G. Wang, and E.~Wilson-Ewing, ``{Bouncing
  cosmologies with dark matter and dark energy},''
  \href{https://dx.doi.org/10.3390/universe3010001}{{\em Universe} {\bfseries
  3} no.~1, (2016) 1}, \href{https://arxiv.org/abs/1610.00938}{{\ttfamily
  arXiv:1610.00938 [astro-ph.CO]}}.

\bibitem{Cai:2017pga}
Y.~Cai, Y.-T. Wang, J.-Y. Zhao, and Y.-S. Piao, ``{Primordial perturbations
  with pre-inflationary bounce},''
  \href{https://dx.doi.org/10.1103/PhysRevD.97.103535}{{\em Phys. Rev. D}
  {\bfseries 97} no.~10, (2018) 103535},
  \href{https://arxiv.org/abs/1709.07464}{{\ttfamily arXiv:1709.07464
  [astro-ph.CO]}}.

\bibitem{Qiu:2018nle}
T.~Qiu, K.~Tian, and S.~Bu, ``{Perturbations of bounce inflation scenario from
  $f(T)$ modified gravity revisited},''
  \href{https://dx.doi.org/10.1140/epjc/s10052-019-6782-8}{{\em Eur. Phys. J.
  C} {\bfseries 79} no.~3, (2019) 261},
  \href{https://arxiv.org/abs/1810.04436}{{\ttfamily arXiv:1810.04436
  [gr-qc]}}.

\bibitem{Hu:2023ndc}
K.~Hu, T.~Paul, and T.~Qiu, ``{Tensor perturbations from bounce inflation
  scenario in f(Q) gravity},''
  \href{https://dx.doi.org/10.1007/s11433-023-2275-0}{{\em Sci. China Phys.
  Mech. Astron.} {\bfseries 67} no.~2, (2024) 220413},
  \href{https://arxiv.org/abs/2308.00647}{{\ttfamily arXiv:2308.00647
  [hep-th]}}.

\bibitem{Zhu:2023lbf}
M.~Zhu, G.~Ye, and Y.~Cai, ``{Pulsar timing array observations as a possible
  hint for nonsingular cosmology},''
  \href{https://arxiv.org/abs/2307.16211}{{\ttfamily arXiv:2307.16211
  [astro-ph.CO]}}.

\bibitem{Ijjas:2020dws}
A.~Ijjas, W.~G. Cook, F.~Pretorius, P.~J. Steinhardt, and E.~Y. Davies,
  ``{Robustness of slow contraction to cosmic initial conditions},''
  \href{https://dx.doi.org/10.1088/1475-7516/2020/08/030}{{\em JCAP} {\bfseries
  08} (2020) 030}, \href{https://arxiv.org/abs/2006.04999}{{\ttfamily
  arXiv:2006.04999 [gr-qc]}}.

\bibitem{Qiu:2013eoa}
T.~Qiu, X.~Gao, and E.~N. Saridakis, ``{Towards anisotropy-free and nonsingular
  bounce cosmology with scale-invariant perturbations},''
  \href{https://dx.doi.org/10.1103/PhysRevD.88.043525}{{\em Phys. Rev. D}
  {\bfseries 88} no.~4, (2013) 043525},
  \href{https://arxiv.org/abs/1303.2372}{{\ttfamily arXiv:1303.2372
  [astro-ph.CO]}}.

\bibitem{Arkani-Hamed:2003pdi}
N.~Arkani-Hamed, H.-C. Cheng, M.~A. Luty, and S.~Mukohyama, ``{Ghost
  condensation and a consistent infrared modification of gravity},''
  \href{https://dx.doi.org/10.1088/1126-6708/2004/05/074}{{\em JHEP} {\bfseries
  05} (2004) 074}, \href{https://arxiv.org/abs/hep-th/0312099}{{\ttfamily
  arXiv:hep-th/0312099}}.

\bibitem{Libanov:2016kfc}
M.~Libanov, S.~Mironov, and V.~Rubakov, ``{Generalized Galileons: instabilities
  of bouncing and Genesis cosmologies and modified Genesis},''
  \href{https://dx.doi.org/10.1088/1475-7516/2016/08/037}{{\em JCAP} {\bfseries
  08} (2016) 037}, \href{https://arxiv.org/abs/1605.05992}{{\ttfamily
  arXiv:1605.05992 [hep-th]}}.

\bibitem{Kobayashi:2016xpl}
T.~Kobayashi, ``{Generic instabilities of nonsingular cosmologies in Horndeski
  theory: A no-go theorem},''
  \href{https://dx.doi.org/10.1103/PhysRevD.94.043511}{{\em Phys. Rev. D}
  {\bfseries 94} no.~4, (2016) 043511},
  \href{https://arxiv.org/abs/1606.05831}{{\ttfamily arXiv:1606.05831
  [hep-th]}}.

\bibitem{Akama:2017jsa}
S.~Akama and T.~Kobayashi, ``{Generalized multi-Galileons, covariantized new
  terms, and the no-go theorem for nonsingular cosmologies},''
  \href{https://dx.doi.org/10.1103/PhysRevD.95.064011}{{\em Phys. Rev. D}
  {\bfseries 95} no.~6, (2017) 064011},
  \href{https://arxiv.org/abs/1701.02926}{{\ttfamily arXiv:1701.02926
  [hep-th]}}.

\bibitem{Cai:2016thi}
Y.~Cai, Y.~Wan, H.-G. Li, T.~Qiu, and Y.-S. Piao, ``{The Effective Field Theory
  of nonsingular cosmology},''
  \href{https://dx.doi.org/10.1007/JHEP01(2017)090}{{\em JHEP} {\bfseries 01}
  (2017) 090}, \href{https://arxiv.org/abs/1610.03400}{{\ttfamily
  arXiv:1610.03400 [gr-qc]}}.

\bibitem{Cai:2017tku}
Y.~Cai, H.-G. Li, T.~Qiu, and Y.-S. Piao, ``{The Effective Field Theory of
  nonsingular cosmology: II},''
  \href{https://dx.doi.org/10.1140/epjc/s10052-017-4938-y}{{\em Eur. Phys. J.
  C} {\bfseries 77} no.~6, (2017) 369},
  \href{https://arxiv.org/abs/1701.04330}{{\ttfamily arXiv:1701.04330
  [gr-qc]}}.

\bibitem{Cai:2017dyi}
Y.~Cai and Y.-S. Piao, ``{A covariant Lagrangian for stable nonsingular
  bounce},'' \href{https://dx.doi.org/10.1007/JHEP09(2017)027}{{\em JHEP}
  {\bfseries 09} (2017) 027},
  \href{https://arxiv.org/abs/1705.03401}{{\ttfamily arXiv:1705.03401
  [gr-qc]}}.

\bibitem{Li:2018ixg}
C.~Li, Y.~Cai, Y.-F. Cai, and E.~N. Saridakis, ``{The effective field theory
  approach of teleparallel gravity, $f(T)$ gravity and beyond},''
  \href{https://dx.doi.org/10.1088/1475-7516/2018/10/001}{{\em JCAP} {\bfseries
  10} (2018) 001}, \href{https://arxiv.org/abs/1803.09818}{{\ttfamily
  arXiv:1803.09818 [gr-qc]}}.

\bibitem{Kolevatov:2017voe}
R.~Kolevatov, S.~Mironov, N.~Sukhov, and V.~Volkova, ``{Cosmological bounce and
  Genesis beyond Horndeski},''
  \href{https://dx.doi.org/10.1088/1475-7516/2017/08/038}{{\em JCAP} {\bfseries
  08} (2017) 038}, \href{https://arxiv.org/abs/1705.06626}{{\ttfamily
  arXiv:1705.06626 [hep-th]}}.

\bibitem{Mironov:2018oec}
S.~Mironov, V.~Rubakov, and V.~Volkova, ``{Bounce beyond Horndeski with GR
  asymptotics and $\gamma$-crossing},''
  \href{https://dx.doi.org/10.1088/1475-7516/2018/10/050}{{\em JCAP} {\bfseries
  10} (2018) 050}, \href{https://arxiv.org/abs/1807.08361}{{\ttfamily
  arXiv:1807.08361 [hep-th]}}.

\bibitem{Boruah:2018pvq}
S.~S. Boruah, H.~J. Kim, M.~Rouben, and G.~Geshnizjani, ``{Cuscuton bounce},''
  \href{https://dx.doi.org/10.1088/1475-7516/2018/08/031}{{\em JCAP} {\bfseries
  08} (2018) 031}, \href{https://arxiv.org/abs/1802.06818}{{\ttfamily
  arXiv:1802.06818 [gr-qc]}}.

\bibitem{Banerjee:2018svi}
S.~Banerjee, Y.-F. Cai, and E.~N. Saridakis, ``{Evading the theoretical no-go
  theorem for nonsingular bounces in Horndeski/Galileon cosmology},''
  \href{https://dx.doi.org/10.1088/1361-6382/ab256a}{{\em Class. Quant. Grav.}
  {\bfseries 36} no.~13, (2019) 135009},
  \href{https://arxiv.org/abs/1808.01170}{{\ttfamily arXiv:1808.01170
  [gr-qc]}}.

\bibitem{Volkova:2019jlj}
V.~E. Volkova, S.~A. Mironov, and V.~A. Rubakov, ``{Cosmological Scenarios with
  Bounce and Genesis in Horndeski Theory and Beyond},''
  \href{https://dx.doi.org/10.1134/S1063776119100236}{{\em J. Exp. Theor.
  Phys.} {\bfseries 129} no.~4, (2019) 553--565}.

\bibitem{Mironov:2022quk}
S.~Mironov and A.~Shtennikova, ``{Stable cosmological solutions in Horndeski
  theory},'' \href{https://dx.doi.org/10.1088/1475-7516/2023/06/037}{{\em JCAP}
  {\bfseries 06} (2023) 037},
  \href{https://arxiv.org/abs/2212.03285}{{\ttfamily arXiv:2212.03285
  [gr-qc]}}.

\bibitem{Mironov:2022ffa}
S.~Mironov and V.~Volkova, ``{Stable nonsingular cosmologies in beyond
  Horndeski theory and disformal transformations},''
  \href{https://dx.doi.org/10.1142/S0217751X22500889}{{\em Int. J. Mod. Phys.
  A} {\bfseries 37} no.~14, (2022) 2250088},
  \href{https://arxiv.org/abs/2204.05889}{{\ttfamily arXiv:2204.05889
  [hep-th]}}.

\bibitem{Ganz:2022zgs}
A.~Ganz, P.~Martens, S.~Mukohyama, and R.~Namba, ``{Bouncing cosmology in
  VCDM},'' \href{https://dx.doi.org/10.1088/1475-7516/2023/04/060}{{\em JCAP}
  {\bfseries 04} (2023) 060},
  \href{https://arxiv.org/abs/2212.13561}{{\ttfamily arXiv:2212.13561
  [gr-qc]}}.

\bibitem{Ilyas:2020qja}
A.~Ilyas, M.~Zhu, Y.~Zheng, Y.-F. Cai, and E.~N. Saridakis, ``{DHOST Bounce},''
  \href{https://dx.doi.org/10.1088/1475-7516/2020/09/002}{{\em JCAP} {\bfseries
  09} (2020) 002}, \href{https://arxiv.org/abs/2002.08269}{{\ttfamily
  arXiv:2002.08269 [gr-qc]}}.

\bibitem{Ilyas:2020zcb}
A.~Ilyas, M.~Zhu, Y.~Zheng, and Y.-F. Cai, ``{Emergent Universe and Genesis
  from the DHOST Cosmology},''
  \href{https://dx.doi.org/10.1007/JHEP01(2021)141}{{\em JHEP} {\bfseries 01}
  (2021) 141}, \href{https://arxiv.org/abs/2009.10351}{{\ttfamily
  arXiv:2009.10351 [gr-qc]}}.

\bibitem{Zhu:2021ggm}
M.~Zhu and Y.~Zheng, ``{Improved DHOST Genesis},''
  \href{https://dx.doi.org/10.1007/JHEP11(2021)163}{{\em JHEP} {\bfseries 11}
  (2021) 163}, \href{https://arxiv.org/abs/2109.05277}{{\ttfamily
  arXiv:2109.05277 [gr-qc]}}.

\bibitem{Zhu:2021whu}
M.~Zhu, A.~Ilyas, Y.~Zheng, Y.-F. Cai, and E.~N. Saridakis, ``{Scalar and
  tensor perturbations in DHOST bounce cosmology},''
  \href{https://dx.doi.org/10.1088/1475-7516/2021/11/045}{{\em JCAP} {\bfseries
  11} no.~11, (2021) 045}, \href{https://arxiv.org/abs/2108.01339}{{\ttfamily
  arXiv:2108.01339 [gr-qc]}}.

\bibitem{Creminelli:2004jg}
P.~Creminelli, A.~Nicolis, and M.~Zaldarriaga, ``{Perturbations in bouncing
  cosmologies: Dynamical attractor versus scale invariance},''
  \href{https://dx.doi.org/10.1103/PhysRevD.71.063505}{{\em Phys. Rev. D}
  {\bfseries 71} (2005) 063505},
  \href{https://arxiv.org/abs/hep-th/0411270}{{\ttfamily
  arXiv:hep-th/0411270}}.

\bibitem{Piao:2004jg}
Y.-S. Piao and Y.-Z. Zhang, ``{The Primordial perturbation spectrum from
  various expanding and contracting phases},''
  \href{https://dx.doi.org/10.1103/PhysRevD.70.043516}{{\em Phys. Rev. D}
  {\bfseries 70} (2004) 043516},
  \href{https://arxiv.org/abs/astro-ph/0403671}{{\ttfamily
  arXiv:astro-ph/0403671}}.

\bibitem{Piao:2004uq}
Y.-S. Piao, ``{On the dualities of primordial perturbation spectrums},''
  \href{https://dx.doi.org/10.1016/j.physletb.2004.12.005}{{\em Phys. Lett. B}
  {\bfseries 606} (2005) 245--250},
  \href{https://arxiv.org/abs/hep-th/0404002}{{\ttfamily
  arXiv:hep-th/0404002}}.

\bibitem{Qiu:2012ia}
T.~Qiu, ``{Reconstruction of a Nonminimal Coupling Theory with Scale-invariant
  Power Spectrum},''
  \href{https://dx.doi.org/10.1088/1475-7516/2012/06/041}{{\em JCAP} {\bfseries
  06} (2012) 041}, \href{https://arxiv.org/abs/1204.0189}{{\ttfamily
  arXiv:1204.0189 [hep-ph]}}.

\bibitem{Qiu:2012np}
T.~Qiu, ``{Reconstruction of f(R) models with Scale-invariant Power
  Spectrum},'' \href{https://dx.doi.org/10.1016/j.physletb.2012.10.045}{{\em
  Phys. Lett. B} {\bfseries 718} (2012) 475--481},
  \href{https://arxiv.org/abs/1208.4759}{{\ttfamily arXiv:1208.4759
  [astro-ph.CO]}}.

\bibitem{Shi:2021tmn}
J.~Shi, Z.~Fang, and T.~Qiu, ``{Adiabatic duality: Duality of cosmological
  models with varying slow-roll parameter and sound speed},''
  \href{https://dx.doi.org/10.1103/PhysRevD.104.063520}{{\em Phys. Rev. D}
  {\bfseries 104} no.~6, (2021) 063520},
  \href{https://arxiv.org/abs/2103.11634}{{\ttfamily arXiv:2103.11634
  [gr-qc]}}.

\bibitem{Lyth:2001nq}
D.~H. Lyth and D.~Wands, ``{Generating the curvature perturbation without an
  inflaton},'' \href{https://dx.doi.org/10.1016/S0370-2693(01)01366-1}{{\em
  Phys. Lett. B} {\bfseries 524} (2002) 5--14},
  \href{https://arxiv.org/abs/hep-ph/0110002}{{\ttfamily
  arXiv:hep-ph/0110002}}.

\bibitem{Lyth:2002my}
D.~H. Lyth, C.~Ungarelli, and D.~Wands, ``{The Primordial density perturbation
  in the curvaton scenario},''
  \href{https://dx.doi.org/10.1103/PhysRevD.67.023503}{{\em Phys. Rev. D}
  {\bfseries 67} (2003) 023503},
  \href{https://arxiv.org/abs/astro-ph/0208055}{{\ttfamily
  arXiv:astro-ph/0208055}}.

\bibitem{Libanov:2011bk}
M.~Libanov, S.~Mironov, and V.~Rubakov, ``{Non-Gaussianity of scalar
  perturbations generated by conformal mechanisms},''
  \href{https://dx.doi.org/10.1103/PhysRevD.84.083502}{{\em Phys. Rev. D}
  {\bfseries 84} (2011) 083502},
  \href{https://arxiv.org/abs/1105.6230}{{\ttfamily arXiv:1105.6230
  [astro-ph.CO]}}.

\bibitem{Hinterbichler:2011qk}
K.~Hinterbichler and J.~Khoury, ``{The Pseudo-Conformal Universe: Scale
  Invariance from Spontaneous Breaking of Conformal Symmetry},''
  \href{https://dx.doi.org/10.1088/1475-7516/2012/04/023}{{\em JCAP} {\bfseries
  04} (2012) 023}, \href{https://arxiv.org/abs/1106.1428}{{\ttfamily
  arXiv:1106.1428 [hep-th]}}.

\bibitem{Creminelli:2011mw}
P.~Creminelli, ``{Conformal invariance of scalar perturbations in inflation},''
  \href{https://dx.doi.org/10.1103/PhysRevD.85.041302}{{\em Phys. Rev. D}
  {\bfseries 85} (2012) 041302},
  \href{https://arxiv.org/abs/1108.0874}{{\ttfamily arXiv:1108.0874 [hep-th]}}.

\bibitem{Wang:2011dt}
H.~Wang, T.~Qiu, and Y.-S. Piao, ``{G-Curvaton},''
  \href{https://dx.doi.org/10.1016/j.physletb.2011.12.016}{{\em Phys. Lett. B}
  {\bfseries 707} (2012) 11--21},
  \href{https://arxiv.org/abs/1110.1795}{{\ttfamily arXiv:1110.1795 [hep-ph]}}.

\bibitem{Feng:2013pba}
K.~Feng, T.~Qiu, and Y.-S. Piao, ``{Curvaton with nonminimal derivative
  coupling to gravity},''
  \href{https://dx.doi.org/10.1016/j.physletb.2014.01.008}{{\em Phys. Lett. B}
  {\bfseries 729} (2014) 99--107},
  \href{https://arxiv.org/abs/1307.7864}{{\ttfamily arXiv:1307.7864 [hep-th]}}.

\bibitem{Feng:2014tka}
K.~Feng and T.~Qiu, ``{Curvaton with nonminimal derivative coupling to gravity:
  Full perturbation analysis},''
  \href{https://dx.doi.org/10.1103/PhysRevD.90.123508}{{\em Phys. Rev. D}
  {\bfseries 90} no.~12, (2014) 123508},
  \href{https://arxiv.org/abs/1409.2949}{{\ttfamily arXiv:1409.2949 [hep-th]}}.

\bibitem{Qiu:2016mrx}
T.~Qiu and K.~Feng, ``{Reheating mechanism of the curvaton with nonminimal
  derivative coupling to gravity},''
  \href{https://dx.doi.org/10.1140/epjc/s10052-017-5275-x}{{\em Eur. Phys. J.
  C} {\bfseries 77} no.~10, (2017) 687},
  \href{https://arxiv.org/abs/1608.01752}{{\ttfamily arXiv:1608.01752
  [hep-ph]}}.

\bibitem{Zhang:2022bde}
X.-z. Zhang, L.-h. Liu, and T.~Qiu, ``{Mimetic curvaton},''
  \href{https://dx.doi.org/10.1103/PhysRevD.107.043510}{{\em Phys. Rev. D}
  {\bfseries 107} no.~4, (2023) 043510},
  \href{https://arxiv.org/abs/2207.07873}{{\ttfamily arXiv:2207.07873
  [hep-th]}}.

\bibitem{Xiong:2024vms}
A.~Xiong, X.-z. Zhang, and T.~Qiu, ``{Perturbations of Mimetic Curvaton},''
  \href{https://arxiv.org/abs/2405.06921}{{\ttfamily arXiv:2405.06921
  [gr-qc]}}.

\bibitem{Brans:1961sx}
C.~Brans and R.~H. Dicke, ``{Mach's principle and a relativistic theory of
  gravitation},'' \href{https://dx.doi.org/10.1103/PhysRev.124.925}{{\em Phys.
  Rev.} {\bfseries 124} (1961) 925--935}.

\bibitem{Xue:2011nw}
B.~Xue and P.~J. Steinhardt, ``{Evolution of curvature and anisotropy near a
  nonsingular bounce},''
  \href{https://dx.doi.org/10.1103/PhysRevD.84.083520}{{\em Phys. Rev. D}
  {\bfseries 84} (2011) 083520},
  \href{https://arxiv.org/abs/1106.1416}{{\ttfamily arXiv:1106.1416 [hep-th]}}.

\bibitem{Cai:2007zv}
Y.-F. Cai, T.~Qiu, R.~Brandenberger, Y.-S. Piao, and X.~Zhang, ``{On
  Perturbations of Quintom Bounce},''
  \href{https://dx.doi.org/10.1088/1475-7516/2008/03/013}{{\em JCAP} {\bfseries
  03} (2008) 013}, \href{https://arxiv.org/abs/0711.2187}{{\ttfamily
  arXiv:0711.2187 [hep-th]}}.

\bibitem{Cai:2008qw}
Y.-F. Cai, T.-t. Qiu, R.~Brandenberger, and X.-m. Zhang, ``{A Nonsingular
  Cosmology with a Scale-Invariant Spectrum of Cosmological Perturbations from
  Lee-Wick Theory},'' \href{https://dx.doi.org/10.1103/PhysRevD.80.023511}{{\em
  Phys. Rev. D} {\bfseries 80} (2009) 023511},
  \href{https://arxiv.org/abs/0810.4677}{{\ttfamily arXiv:0810.4677 [hep-th]}}.

\bibitem{Easson:2011zy}
D.~A. Easson, I.~Sawicki, and A.~Vikman, ``{G-Bounce},''
  \href{https://dx.doi.org/10.1088/1475-7516/2011/11/021}{{\em JCAP} {\bfseries
  11} (2011) 021}, \href{https://arxiv.org/abs/1109.1047}{{\ttfamily
  arXiv:1109.1047 [hep-th]}}.

\bibitem{Planck:2018jri}
{\bfseries Planck} Collaboration, Y.~Akrami {\em et~al.}, ``{Planck 2018
  results. X. Constraints on inflation},''
  \href{https://dx.doi.org/10.1051/0004-6361/201833887}{{\em Astron.
  Astrophys.} {\bfseries 641} (2020) A10},
  \href{https://arxiv.org/abs/1807.06211}{{\ttfamily arXiv:1807.06211
  [astro-ph.CO]}}.

\bibitem{Bird:2010mp}
S.~Bird, H.~V. Peiris, M.~Viel, and L.~Verde, ``{Minimally Parametric Power
  Spectrum Reconstruction from the Lyman-alpha Forest},''
  \href{https://dx.doi.org/10.1111/j.1365-2966.2011.18245.x}{{\em Mon. Not.
  Roy. Astron. Soc.} {\bfseries 413} (2011) 1717--1728},
  \href{https://arxiv.org/abs/1010.1519}{{\ttfamily arXiv:1010.1519
  [astro-ph.CO]}}.

\bibitem{Fixsen:1996nj}
D.~J. Fixsen, E.~S. Cheng, J.~M. Gales, J.~C. Mather, R.~A. Shafer, and E.~L.
  Wright, ``{The Cosmic Microwave Background spectrum from the full COBE FIRAS
  data set},'' \href{https://dx.doi.org/10.1086/178173}{{\em Astrophys. J.}
  {\bfseries 473} (1996) 576},
  \href{https://arxiv.org/abs/astro-ph/9605054}{{\ttfamily
  arXiv:astro-ph/9605054}}.

\bibitem{Chluba:2012we}
J.~Chluba, A.~L. Erickcek, and I.~Ben-Dayan, ``{Probing the inflaton:
  Small-scale power spectrum constraints from measurements of the CMB energy
  spectrum},'' \href{https://dx.doi.org/10.1088/0004-637X/758/2/76}{{\em
  Astrophys. J.} {\bfseries 758} (2012) 76},
  \href{https://arxiv.org/abs/1203.2681}{{\ttfamily arXiv:1203.2681
  [astro-ph.CO]}}.

\bibitem{Chluba:2012gq}
J.~Chluba, R.~Khatri, and R.~A. Sunyaev, ``{CMB at 2x2 order: The dissipation
  of primordial acoustic waves and the observable part of the associated energy
  release},'' \href{https://dx.doi.org/10.1111/j.1365-2966.2012.21474.x}{{\em
  Mon. Not. Roy. Astron. Soc.} {\bfseries 425} (2012) 1129--1169},
  \href{https://arxiv.org/abs/1202.0057}{{\ttfamily arXiv:1202.0057
  [astro-ph.CO]}}.

\bibitem{Cyr:2023pgw}
B.~Cyr, T.~Kite, J.~Chluba, J.~C. Hill, D.~Jeong, S.~K. Acharya, B.~Bolliet,
  and S.~P. Patil, ``{Disentangling the primordial nature of stochastic
  gravitational wave backgrounds with CMB spectral distortions},''
  \href{https://dx.doi.org/10.1093/mnras/stad3861}{{\em Mon. Not. Roy. Astron.
  Soc.} {\bfseries 528} no.~1, (2024) 883--897},
  \href{https://arxiv.org/abs/2309.02366}{{\ttfamily arXiv:2309.02366
  [astro-ph.CO]}}.

\bibitem{Byrnes:2018txb}
C.~T. Byrnes, P.~S. Cole, and S.~P. Patil, ``{Steepest growth of the power
  spectrum and primordial black holes},''
  \href{https://dx.doi.org/10.1088/1475-7516/2019/06/028}{{\em JCAP} {\bfseries
  06} (2019) 028}, \href{https://arxiv.org/abs/1811.11158}{{\ttfamily
  arXiv:1811.11158 [astro-ph.CO]}}.

\bibitem{Chen:2022usd}
J.-W. Chen, M.~Zhu, S.-F. Yan, Q.-Q. Wang, and Y.-F. Cai, ``{Enhance primordial
  black hole abundance through the non-linear processes around bounce point},''
  \href{https://dx.doi.org/10.1088/1475-7516/2023/01/015}{{\em JCAP} {\bfseries
  01} (2023) 015}, \href{https://arxiv.org/abs/2207.14532}{{\ttfamily
  arXiv:2207.14532 [astro-ph.CO]}}.

\bibitem{Papanikolaou:2024fzf}
T.~Papanikolaou, S.~Banerjee, Y.-F. Cai, S.~Capozziello, and E.~N. Saridakis,
  ``{Primordial black holes and induced gravitational waves in non-singular
  matter bouncing cosmology},''
  \href{https://dx.doi.org/10.1088/1475-7516/2024/06/066}{{\em JCAP} {\bfseries
  06} (2024) 066}, \href{https://arxiv.org/abs/2404.03779}{{\ttfamily
  arXiv:2404.03779 [gr-qc]}}.

\bibitem{Sharma:2024whg}
M.~K. Sharma, M.~Sami, and D.~F. Mota, ``{Generic predictions for primordial
  perturbations and their implications},''
  \href{https://dx.doi.org/10.1016/j.physletb.2024.138956}{{\em Phys. Lett. B}
  {\bfseries 856} (2024) 138956},
  \href{https://arxiv.org/abs/2401.11142}{{\ttfamily arXiv:2401.11142
  [astro-ph.CO]}}.

\end{thebibliography}\endgroup
\bibliographystyle{utphys}

\end{document}